\documentclass[trackchanges,12pt]{aastex701}
\usepackage{amssymb}
\usepackage{xcolor}
\usepackage[T1]{fontenc} 
\usepackage{amsmath}
\usepackage{subfig}
\usepackage{graphicx}

\usepackage{amsfonts}
\usepackage{mathrsfs}
\usepackage{bm}

\usepackage{amsmath}

\begin{document}

\title{Implications of Matter--Curvature Coupled Gravity on Chandrasekhar Mass Limit of White Dwarfs}

\author[orcid=0009-0000-3417-9764]{N. Priyobarta}
\affiliation{Department of Mathematics, Birla Institute of Technology and Science, Pilani, Hyderabad Campus, Jawahar Nagar, Kapra Mandal, Medchal District, Telangana-500078, India.}
\email[show]{priyo.naoremcha@gmail.com}  

\author[orcid=0000-0003-4089-3651]{S. K. Maurya}
\affiliation{Department of Mathematical and Physical Sciences,
College of Arts and Sciences, University of Nizwa, P.O. Box 33, Nizwa 616, Sultanate of Oman}
\email[show]{sunil@unizwa.edu.om (corresponding author) }

\author[orcid=0000-0001-9778-4101]{Ksh. Newton Singh} 
\affiliation{Department of Physics \& Astrophysics, University of Delhi, Delhi-110007, India.}
\email[show]{ntnphy@gmail.com}

\author[orcid=0000-0001-5527-3565]{B. Mishra}
\affiliation{Department of Mathematics, Birla Institute of Technology and Science, Pilani, Hyderabad Campus, Jawahar Nagar, Kapra Mandal, Medchal District, Telangana-500078, India.}
\email[show]{bivu@hyderabad.bits-pilani.ac.in}

\begin{abstract}

We investigate white dwarfs in the framework of $f(\mathcal{R,L}_m)$ and $f(\mathcal{R,L}_m,\mathcal{T})$ gravity to explore the Chandrasekhar Limit. We have considered two functional forms of $f(\mathcal{R,L}_m)$ and one functional form of $f(\mathcal{R,L}_m,\mathcal{T})$ gravity. Considering the matter Lagrangian $\mathcal{L}_m=p$, we calculate modified TOV equations for each of the forms. By employing the fully degenerate electron gas equation of state in the modified TOV equations, we derive the mass-radius relation for each functional form of both $f(\mathcal{R,L}_m)$ and $f(\mathcal{R,L}_m, \mathcal{T})$ gravity. Our models imply modifications in the Chandrasekhar mass limit that deviate significantly from the GR and the Newtonian cases. In the $f(\mathcal{R,L}_m, \mathcal{T})$ gravity, the new mass limit of the white dwarf can reach {up to} $1.537\,\mathrm{M}_\odot$ while in $f(\mathcal{R,L}_m)$ with the quadratic extension can reach up to $1.52\,\mathrm{M}_\odot$ and with square-root exponential model extension {up to} $2.08\,\mathrm{M}_\odot$. Further, we analyze the static stability criterion, the gravitational redshift, and the adiabatic indices. For the power-law form of $f(\mathcal{R,L}_m)$ and the non-linear form of  $f(\mathcal{R,L}_m,\mathcal{T})$ gravity, significant variations are observed at higher densities ($\rho_c > 10^{10}\, \mathrm{g/cm^3}$), while substantial changes are noted at much lower central densities in the case of square-root exponential model form of $f(\mathcal{R,L}_m)$ gravity. {The higher-density configurations should be regarded as formal equilibrium solutions within the adopted Chandrasekhar EoS and not as fully realistic stable C--O WDs, since additional high-density microphysics has not been included. We also calculate compactness and gravitational redshift, which are much lower than those of NS and BH. The configuration satisfies the adiabatic stability criterion, which show that all considered models yield $\Gamma > 4/3$ throughout the interiors of WDs. Overall, our models provide a viable framework for the existence of super-Chandrasekhar mass limit, extending beyond the classical predictions in the Newtonian and/or GR cases.}

\end{abstract}

\keywords{White Dwarfs,~~$f(\mathcal{R,L}_m)$ gravity,~~ $f(\mathcal{R,L}_m,\mathcal{T})$ gravity,~~ Chandrasekhar's equation of state,~~Mass--Radius Relation,~~Stability Analysis. }


\section{Introduction}\label{Sec:I}
The general theory of relativity (GR) provides a fundamental theoretical framework for understanding various cosmic phenomena. Since its inception, GR has been crucial in advancing our understanding of a range of astrophysical and cosmological phenomena, including the behavior of planetary systems and the evolution of the Universe. Throughout the last centuries, this theory has been rigorously tested through both experiments and observations, demonstrating outstanding consistency with observed data in both weak-field and strong-field conditions. The key validation of GR includes the precise prediction of the precession of Mercury’s orbit \citep{Einstein_1915}, the detection of gravitational waves (GWs) resulting from the collisions of binary black holes and neutron stars via the LIGO and Virgo observatories  \citep{Abbott_2016,Abbott_2017_18} and the observation of black hole's shadow by the Event Horizon Telescope  \citep{Akiyama_2019}. These findings strongly support the predictions made by GR, relying on data primarily gathered since the early twentieth century.  Nevertheless, GR encounters with challenges when addressing certain recent astronomical discoveries. Observations of distant supernovae in the late 1990s revealed that the universe is expanding at an accelerating pace  \citep{Riess_1998,Perlmutter_1999,Knop_2003,Amanullah_2010,Weinberg_2013}, a behavior that GR cannot account for without incorporating additional concepts, such as dark energy. Furthermore, in the field of astrophysical structures, various researchers have formulated theories to explain compact objects, such as neutron stars (NS), white dwarfs (WDs) and black holes (BHs), which are in the strong gravity regime.

One of the remarkable stellar objects--the WD, which forms when a star has depleted its nuclear fuel and retains a compact core primarily composed of helium, carbon, and oxygen  \citep{Shapiro_1983,Lauffer_2018}. More than $95\%$ of all the observed stars in the Milky Way are WDs \citep{Woosley_2015}. In most cases, WDs have carbon-oxygen (C-O) cores, which allow for larger masses compared to those of helium-core WDs. WDs resist gravitational collapse due to electron degeneracy pressure, a phenomenon arising from the Pauli exclusion principle in quantum mechanics that ultimately establishes a maximum stable mass. Chandrasekhar precisely predicted the maximum mass of the WDs for are non-rotating and non-magnetized as $1.4\,\mathrm{M_\odot}$  \citep{Chandrasekhar_1931,Chandrasekhar_1931_456}. This is the upper mass limit for WDs known as ``the Chandrasekhar mass limit''. Recent research has revealed the existence of both over-luminous \citep{andrew_2006,scalzo_2010,yamanaka_2009,Silverman_2013} and under-luminous \citep{Filippenko_1992,Taubenberger_2008,Turatto_1998} type Ia supernovae (SNe Ia), suggesting that the Chandrasekhar limit may not be universally applicable. Additionally, various observations indicate that there are masses in the range of $2.1-2.8 \,\mathrm{M}_\odot$, which significantly surpass the upper limit \citep{Hicken_2007,khokhlov_1993}, thereby challenging the uniqueness of the Chandrasekhar limit and the current theory of GR.

These observations, along with the cosmological problems, motivate us to modify the existing theory of GR \citep{Buchdahl_1970,Capozziello_2011}. By treating the Lagrangian as an arbitrary function of the Ricci scalar $\mathcal{R}$, \cite{Felice_2010} develop a class of $f(\mathcal{R})$ theories that serve as an alternative explanation for the accelerated expansion of the universe, without considering dark energy. On top of that, $f(\mathcal{R})$ models provide a wide variety of phenomenological implications in cosmological \citep{Capozziello_2006_135,Amendola_2008,Tsujikawa_2008,Liu_2018}.  In the field of compact objects like NS, the $f(\mathcal{R})$ theories provided the very prominent results \citep{Astashenok_2013,Nashed_2022,Nashed_2021,Astashenok_2020_135,Capozziello_2015,Astashenok_2021}. \cite{Ganguly_2014} formulated NS with the functional form of $f(\mathcal{R}) = \mathcal{R}+ \alpha \mathcal{R}^2$, which is also known as the Starobinsky model, and described how some of the EoS can match all conditions provided at the surface of the NS. \cite{Astashenok_2017} provided the stable NS of maximum mass $> 2\, \mathrm{M}_\odot$  using various relativistic EoS. Further, by considering $f(\mathcal{R})= \mathcal{R}+ \mathcal{R}\left[ \exp (\mathcal{-R}/\mathcal{R}_0)-1 \right]$ and $\mathcal{R}^2$ models with logarithmic and cubic corrections, they have obtained $2.2 -2.3\, \mathrm{M}_\odot$ \citep{Astashenok_2015}, which explains the mass-gap of the NS-BHs in astrophysical scenarios.  In the context of WDs, many researchers have also been interested in studying their mass and radius in light of modified theories \citep{das_2015,banerjee_2017,Kalita_2018,Jain_2016,Astashenok_2022}.  In addition to the $f(\mathcal{R})$ theories, various modified gravity theories are formulated and analyzed WDs such as $f(\mathcal{R,T})$ \citep{Carvalho_2017,Rocha_2020}, Palatini $f(\mathcal{R})$ \citep{Kalita_2022,Kalita_2023}, Rastall-Rainbow \citep{Li_2023} and scalar-tensor theories \citep{Vidal_2024}.

One way to modify the theory is by considering non-minimal matter--curvature coupling to gravity. In this theory, matter and geometry are non-minimally coupled to each other \citep{Bertolami_2009,Harko_2010_044,Lobo_2025}. 
As a similar approach to $f(\mathcal{R})$ gravity,  \cite{Harko_2010_1} proposed the $f(\mathcal{R,L}_m)$ gravity theory, which generalizes the action to an arbitrary function that depends on both $\mathcal{R}$ and $\mathcal{L}_m$. This theory can be viewed as an extensive expansion of earlier models, merging contributions from both curvature and matter into a unified functional expression. The non-minimal coupling present in $f(\mathcal{R,L}_m)$ gravity indicates a covariant non-conservation of the energy-momentum tensor, resulting in the non-geodesic motion of test particles and the emergence of an additional force. Notably, the explicit violation of the equivalence principle renders this theory amenable to direct testing through astrophysical observations and solar-system constraints \citep{Faraoni_2009,Bertolami_2006}.

In the context of $f(\mathcal{R}, \mathcal{L}_m)$ gravity, researchers have examined various phenomena, particularly compact stars. \cite{Lobato_2021,lobato_2022} investigated the interaction between $\mathcal{R}$ and $\mathcal{L}_m$, which constrain GWs and massive pulsars such as GW170817 and PSR J0030+0451. Additionally, studies of wormholes have uncovered intriguing structures within the $f(\mathcal{R,L}_m)$ gravity framework, yielding positive results \citep{Lobo_2009, Agrawal_2025, Khatri_2024}. Recent research has also looked into energy conditions \citep{Wang_2012} and the cosmological effects \citep{Nesseris_2009, Goncalves_2023, Harko_2010_8} related to $f(\mathcal{R,L}_m)$ gravity. \cite{lobato_2022} considered $f(\mathcal{R,L}_m) = \mathcal{R}/16\pi + \mathcal{L}_m + \sigma \mathcal{RL}_m$ predicted the maximum masses which are above the Chandrasekhar mass limit.

Furthermore, building upon the concepts of unification of  $f(\mathcal{R,L}_m)$ \citep{Harko_2010_1} and $f(\mathcal{R,T})$ \citep{Harko_2011} gravity, \cite{Haghani_2021_81} developed these theories into a more generalized coupling between geometry and matter, referred to as $f(\mathcal{R,L}_m,\mathcal{T})$ gravity. In this theory, the Lagrangian is considered as an arbitrary function of the Ricci scalar $\mathcal{R}$, the matter Lagrangian $\mathcal{L}_m$, and the trace of the energy-momentum tensor $\mathcal{T}$.  Further, one can still reduce  $f(\mathcal{R,L}_m,\mathcal{T})$ gravity  to $f(\mathcal{R})$, $f(\mathcal{R,L}_m)$ and $f(\mathcal{R,T})$ gravity as provided in Ref. \citep{Priyobarta_2025_162}. Recently, extensive research has explored various aspects of $f(\mathcal{R}, \mathcal{L}_m, \mathcal{T})$ gravity, such as wormholes \citep{Moraes_2024}, compact objects \citep{Priyobarta_2025_162, Mota_2024}, and in the context of cosmological implications \citep{kshirsagar_2025}. In the context of $f(\mathcal{R,L}_m,\mathcal{T})$, \cite{Otoniel_2026} provided the possibility of the existence of WDs beyond the Chandrasekhar mass limit. 

In this article, we will investigate both $f(\mathcal{R,L}_m)$ and $f(\mathcal{R,L}_m,\mathcal{T})$ gravity. We will study their impacts on the structure of WDs, focusing on mass-radius relations and the stability of the system. For $f(\mathcal{R,L}_m)$ gravity,  \cite{Taser_2025}  considered the power law and investigated the Krori-Barua spacetime and constructed analytical compact star solutions specially for NS and also motivated by the square-root exponential form proposed by \cite{Cai_2020} in the context of cosmology within modified gravity in the framework of $f(\mathcal{R,L}_m)$ we will consider two functional models i.e., power-law $\big[f(\mathcal{R,L}_m) = {\mathcal{R} / 16 \pi} +\mathcal{L}_m + \sigma \mathcal{L}^{2}_m\big]$ and  square-root exponential form $\left[f(\mathcal{R,L}_m) = {\mathcal{R}/ 16 \pi} +\Big[1-\gamma  \exp \Big(-\sqrt{\beta  \mathcal{L}_m}\Big)\Big] \mathcal{L}_m \right]$, whereas in the case of $f(\mathcal{R,L}_m,\mathcal{T})$, a non-linear functional form $\left[f(\mathcal{R},\mathcal{L}_m,\mathcal{T}) = \mathcal{R}+\alpha \mathcal{L}_m \mathcal{T} \right]$ are considered. We also analyze various stability criteria in order to ensure the physical viability of the WD models. Throughout the article, we will adopt the system of units in which $G = c = 1$. 

The structure of the article: in section \ref{Sec:II}, the field equations for both $f(\mathcal{R,L}_m)$ and $f(\mathcal{R,L}_m,\mathcal{T})$ gravity are introduced, along with their respective models. Section \ref{Sec:III} includes the computation of the TOV equations for each model, while section \ref{Sec:IV} offers a brief summary of Chandrasekhar's equation of state for degenerate WDs. The numerical analysis and the mass-radius analysis are presented in section \ref{Sec:V}. Various stability analyses are conducted in section \ref{Sec:VI}. Finally, section \ref{Sec:VII} presents the conclusions.

\section{Mathematical Formalism}\label{Sec:II}

\subsection{The $f(\mathcal{R},\mathcal{L}_m)$ gravity}
The theory provides the gravitational Lagrangian as a function of Ricci scalar $\mathcal{R}$ and matter Lagrangian  $\mathcal{L}_m$. In this framework, the action of $f(\mathcal{R,L}_m)$ gravity  \citep{Harko_2010_1} is given as, 
\begin{equation}
S = \int f(\mathcal{R,L}_m) \sqrt{-g} \,d^4x,
\label{eq:frlmaction}
\end{equation}
where $f(\mathcal{R,L}_m)$ be an arbitrary function of $\mathcal{R}$ and $\mathcal{L}_m$, respectively denote the Ricci scalar and matter Lagrangian. Assuming the matter Lagrangian only depends on the metric \citep{Landau_1975}, we defined  the energy-momentum tensor of the matter as 
\begin{equation}
\mathcal{T}_{\mu \nu}= g_{\mu \nu} \,\mathcal{L}_m -2\,{\partial \mathcal{L}_m \over \partial g^{\mu \nu}}~.
\end{equation}
By varying the action \eqref{eq:frlmaction} with respect to metric, the field equations of $f(\mathcal{R,L}_m)$ gravity can be obtained as
\begin{eqnarray}
&&\hspace{-2cm} f_{_\mathcal{R}}(\mathcal{R,L}_m)\, \mathcal{G}_{\mu\nu} 
+ {1 \over 2}\, g_{\mu\nu}\Big[ f_{_\mathcal{R}}(\mathcal{R,L}_m)\,\mathcal{R}-f(\mathcal{R,L}_m) + f_{_\mathcal{L}}(\mathcal{R,L}_m)\,\mathcal{L}_m \Big]   \nonumber \\
&& \hspace{4cm} = {1 \over 2}\,f_{_\mathcal{L}}(\mathcal{R,L}_m) \,\mathcal{T}_{\mu\nu} -\Big(g_{\mu\nu} \square - \nabla_\mu \nabla_\nu \Big)f_{_\mathcal{R}}(\mathcal{R,L}_m)~.
\label{eq:frlm}
\end{eqnarray}
For brevity, we denote 
\begin{eqnarray*}
f_{_\mathcal{R}}(\mathcal{R,L}_m) = {\partial f(\mathcal{R,L}_m) \over \partial \mathcal{R}} ~~~\mbox{and} ~~~
f_{_\mathcal{L}}(\mathcal{R,L}_m) = {\partial f(\mathcal{R,L}_m) \over \partial\mathcal{L}_m}~.
\end{eqnarray*} 
On substituting $f(\mathcal{R},\mathcal{L}_m)= \mathcal{R}/{16\pi}+\mathcal{L}_m$ into Eq.~\eqref{eq:frlm}, one recovers the standard field equations of GR, $\mathcal{G}_{\mu\nu}=8\pi \mathcal{T}_{\mu\nu}$. The divergence of the energy-momentum tensor $\mathcal{T}_{\mu\nu}$ can be given as,
\begin{eqnarray}
\nabla^\mu \mathcal{T}_{\mu \nu} = \nabla^\mu \ln \Big[ f_{_\mathcal{L}}(\mathcal{R,L}_m) \Big]\left( \mathcal{L}_m \,g_{\mu\nu} - \mathcal{T}_{\mu\nu} \right).
\label{eq:emfrlm}
\end{eqnarray}
In this article, we have proposed two functional forms of $f(\mathcal{R,L}_m)$ gravity as

\begin{itemize}
\item Power-law  form \citep{Harko_2010_1}:
\begin{eqnarray}
\label{eq:fRLmM1}
f(\mathcal{R,L}_m) = {\mathcal{R} \over 16 \pi} +\mathcal{L}_m + \sigma \mathcal{L}^{2}_m   ~, 
\end{eqnarray}
where $\sigma$ is coupling parameter and has a dimension of $\mathrm{cm^2/dyne}$.

\item Square-root exponential  form \citep{Harko_2010_1}:
\begin{eqnarray}
\label{eq:fRLmM2}
f(\mathcal{R,L}_m) = {\mathcal{R} \over 16 \pi} +\Big[1-\gamma  \exp \Big(-\sqrt{\beta  \mathcal{L}_m}\Big)\Big] \mathcal{L}_m ~,
\end{eqnarray}
where $\gamma$ and $\beta$ are coupling parameter. For simplicity, $\beta$ is fixed at 2, and the values of  $ \gamma$ are varied. The parameter $\beta$ has a unit of $\mathrm{cm^2/dyne}$  while $\gamma$ is dimensionless parameter.
\end{itemize}

\subsection{ The $f(\mathcal{R},\mathcal{L}_m,\mathcal{T})$ gravity}
As an extension to the $f(\mathcal{R},\mathcal{L}_m)$ gravity, we also consider $f(\mathcal{R},\mathcal{L}_m, \mathcal{T})$, which non minimally couples with the trace of the energy-momentum tensor.
The action for $f(\mathcal{R},\mathcal{L}_m,\mathcal{T})$ gravity  \citep{Haghani_2021_81} can be given as,
\begin{eqnarray}
S = {1\over 16 \pi}\int\, f(\mathcal{R},\mathcal{L}_m,\mathcal{T}) \,\sqrt{-g}\,\, d^4x + \int \mathcal{L}_m\,\sqrt{-g}\,\, d^4x \, . \label{eq:action}
\end{eqnarray}
Here, the function $f(\mathcal{R},\mathcal{L}_m,\mathcal{T})$ is an arbitrary function of the Ricci scalar $\mathcal{R}$, matter Lagrangian $\mathcal{L}_m$ and $\mathcal{T}$ be the trace of energy momentum tensor $\mathcal{T}_{\mu \nu}$. Varying the action (\ref{eq:action}) with respect to the metric $g^{\mu \nu}$, the field equations of $f(\mathcal{R},\mathcal{L}_m,\mathcal{T})$ gravity obtained as,
\begin{eqnarray}
\label{eq:fieldeqn}
f_{_\mathcal{R}}\mathcal{R}_{\mu \nu}+ \big(g_{\mu \nu}\square - \nabla_\mu \nabla_\nu \big) f_{_\mathcal{R}}- {1\over2}\Big[ f-(f_{_\mathcal{L}} + 2f_{_\mathcal{T}} )\mathcal{L}_m\Big]g_{\mu \nu} =\Big[8\pi + {1\over 2}(f_{_\mathcal{L}}+ 2f_{_\mathcal{T}} )\Big] \mathcal{T}_{\mu \nu}+ \tau_{\mu \nu} f_{_\mathcal{T}}.
\end{eqnarray}
Here,~$f=f(\mathcal{R},\mathcal{L}_m,\mathcal{T})$, $ f_{_\mathcal{R}}= \partial f /\partial \mathcal{R}$, $f_{_\mathcal{L}}= \partial f /\partial \mathcal{L}_m$, $f_{_\mathcal{T}}= \partial f /\partial \mathcal{T}$, $\mathcal{R}_{\mu \nu}$ represents the Ricci tensor and $\nabla $ be the covariant derivative with respect to the symmetric connection associated to $g_{\mu \nu}$. On substitution, $f(\mathcal{R,L}_m,\mathcal{T})=\mathcal{R}$, Eq.\eqref{eq:fieldeqn} can reduce to GR. The tensor, $\tau_{\mu \nu}$ can be defined as  \citep{Haghani_2021_81},
\begin{eqnarray}
    \label{eq:tau}
    \tau_{\mu \nu} = 2 \,g^{\sigma \delta} {\partial^2 \mathcal{L}_m \over \partial g^{\mu \nu} \,\partial g^{\sigma \delta} } \, .
 \end{eqnarray}
Taking the covariant derivative of Eq. \eqref{eq:fieldeqn} and using the geometric identity only  \citep{Koivisto_2006} $[ \Box, \nabla_\nu]f_{_\mathcal{R}} =\mathcal{R}_{\mu \nu}\nabla^\mu  f_{_\mathcal{R}}$, one can obtain the non-conservative energy-momentum tensor as, 
\begin{eqnarray}\label{eq:consrv}
&& \nabla^\mu \mathcal{T}_{\mu \nu} = {1 \over8 \pi + f_m}\Big[ \nabla_\nu (\mathcal{L}_m f_m) - \mathcal{T}_{\mu \nu} \nabla^\mu  f_m - \mathcal{A}_\nu -{1 \over 2}\Big( f_{_\mathcal{T}} \nabla_\nu \mathcal{T} + f_{_\mathcal{L}} \,\nabla_\nu  \mathcal{L}_m
 \Big)\Big] \, ,~~~~~~~~~
\end{eqnarray}
where $f_m = f_{_\mathcal{T}}+{1 \over 2} f_{_\mathcal{L}} ~~~\mbox{and}~~~ \mathcal{A}_\nu =\nabla^\mu (f_{_\mathcal{T}}\, \tau_{\mu \nu})$. 
In case of $f(\mathcal{R},\mathcal{L}_m, \mathcal{T})$  will be consider only the non-linear or multiplicative case \citep{Haghani_2021_81}:
\begin{eqnarray}
\label{eq:fRLmTM1}
f(\mathcal{R},\mathcal{L}_m,\mathcal{T}) = \mathcal{R} +\alpha \,\mathcal{L}_m \mathcal{T}\,,
\end{eqnarray}
where $\alpha$ is the coupling parameter with the units of $\mathrm{cm^2/dyne}$.

\subsection{Compact Stellar Models}
Throughout the paper, we consider the static spherically symmetric 4-$D$ spacetime as given by the metric,
\begin{eqnarray}\label{metric}
ds^2= - e^{\xi(r)}dt^2 + e^{\lambda(r)}dr^2 + r^{2}d\theta^2 + r^{2} \sin^2\theta \,d \phi^2,  
\end{eqnarray} 
where $\xi(r)$ and $\lambda(r)$ are functions of $r$ only. The interior of the star is an isotropic and perfect fluid, given as
\begin{eqnarray}
\mathcal{T}_{\mu \nu}= (\rho + p)u_\mu u_\nu + p\, g_{\mu \nu},
\label{eq:emtensor}
\end{eqnarray}
where $\rho(r)$ and $p(r)$ are the energy density and pressure of matter, respectively. $u_\mu = \bigl\{e^{\,\xi/2},0,0,0\bigr\}$  denotes the four velocity of the fluid. In this work, we adopt the matter Lagrangian, $\mathcal{L}_m = p$. It is important to note that in theories with explicit matter--curvature coupling, different choices i.e., $\mathcal{L}_m = p$ or $\mathcal{L}_m = -\rho$ are generally not equivalent. These different choices can lead to different field equations and conservation laws. { We have chosen $\mathcal{L}_m = p$, as it simplifies the field equations  and is also widely adopted in cosmological models} \citep{Harko_2010_1}.

\subsubsection{Power-law form  of $f(\mathcal{R,L}_m)$ gravity}
For the power-law choice of $f(\mathcal{R,L}_m)$  gravity \eqref{eq:fRLmM1}, the field equation \eqref{eq:frlm} becomes
\begin{eqnarray}
\mathcal{G}_{\mu\nu} = 8\pi \Big[ \left(1+ 2\sigma \ \mathcal{L}_m \right)\mathcal{T}_{\mu\nu} - \sigma \,g_{\mu\nu} \,\mathcal{L}^{2}_m  \Big],
\end{eqnarray}
where the $tt$ and $rr$ components respectively become,
\begin{eqnarray}
&&\hspace{-1.7cm} e^{-\lambda  } \left({ \lambda '  \over r}- {1\over r^2}\right) +{1 \over r^2} 
= 8 \pi \sigma \,p^{2} +8 \pi \left( 1+2\sigma\, \,p\right) \rho,~~
    \label{eq:fRLM1Gtt}\\
&&\hspace{-1.7cm}e^{ -\lambda}\left( {\xi' \over r} + {1 \over r^2}\right) -{1 \over r^2}= 8 \pi \left(p + \sigma \,p^{2}\right)
    \label{eq:fRLM1Grr}.
\end{eqnarray}
From the covariant derivative of energy-momentum tensor \eqref{eq:emfrlm}, we get
\begin{eqnarray}
    {dp \over dr}= -(\rho +p)\,{\xi' \over 2}\,.
    \label{eq:fRLM1em}
\end{eqnarray}
\subsubsection{Square-root exponential form of $f(\mathcal{R,L}_m)$ gravity}
For the square-root exponential form of $f(\mathcal{R,L}_m)$ gravity \eqref{eq:fRLmM2}, the field equation \eqref{eq:frlm} becomes
\begin{eqnarray}
\mathcal{G}_{\mu \nu}= 8\pi \left[ \frac{1}{2} \gamma  \,e^{-\sqrt{\beta \, p}} \left(\sqrt{\beta  \,p}-2\right)+1 \right] \mathcal{T}_{\mu \nu} - 4 \pi \,g_{\mu \nu} \left[  \gamma \, p\, e^{-\sqrt{\beta \, p}} \sqrt{\beta \, p}\right], 
\end{eqnarray}
with $tt$ and $rr$ components are respectively,
\begin{eqnarray}\label{eq:fRLM2Gtt}
\hspace{-1.2cm} \sqrt{\beta \, p} \left[ e^{\sqrt{\beta \, p}} \left(r \lambda '+e^{\lambda }-1\right)- 8 \pi  r^2 e^{\lambda } \rho  \left(e^{\sqrt{\beta \, p}}-\gamma \right)\right] &=& \, 4 \pi  \gamma\,  \beta  r^2 p\, e^{\lambda } \rho +4 \pi  \gamma\,  \beta  r^2 p^2 \,e^{\lambda }, \\ 
e^{\lambda }-r\, \xi '-1 &=& \,8 \pi  r^2 p \,e^{\lambda } \left(\gamma \, e^{-\sqrt{\beta \,  p}}-1\right)~. \label{eq:fRLM2Grr}
\end{eqnarray}
Also, from Eq. \eqref{eq:emfrlm} we obtained 
\begin{eqnarray}
{dp \over dr}= -(\rho +p)\,{\xi' \over 2}\,.
    \label{eq:fRLM2em}
\end{eqnarray}

\subsubsection{Non-linear Form of $f(\mathcal{R},\mathcal{L}_m,\mathcal{T})$ gravity}
 With the non-linear model \eqref{eq:fRLmTM1}, the field equation \eqref{eq:fieldeqn} reduces to 
\begin{equation}
G_{\mu \nu} = \Big[ 8\pi +{\alpha \over 2}\,(5p - \rho)\Big] \mathcal{T}_{\mu \nu} + \alpha \,p^2 g_{\mu \nu}~.
\end{equation}
The $tt$ and $rr$ components are obtained respectively as,
\begin{eqnarray}
e^{-\lambda  } \left({ \lambda '  \over r}- {1\over r^2}\right) +{1 \over r^2} &=&  \frac{1}{2} \Big[5 \alpha  \,p \,\rho +2 \alpha \, p^2+\rho \,(16 \pi -\alpha \, \rho )\Big] \label{eq:fRLTGtt},\\
e^{ -\lambda}\left( {\xi' \over r} + {1 \over r^2}\right) -{1 \over r^2} &=& 
\frac{1}{2} \Big[ 16 \pi+3 \alpha \, p-\alpha \, \rho  \Big]p  \,.\label{eq:fRLTGrr}
\end{eqnarray}
Also, from the covariant derivative of energy-momentum tensor \eqref{eq:consrv}, we obtain
\begin{equation}
p' +(\rho + p) \,{\xi' \over 2} = \frac{\alpha \, p \left(\rho'- p'\right)}{16 \pi+\alpha(5   p-  \rho)  }~~,
    \label{eq:fRLTem}
\end{equation}
where prime denotes the derivative with respect to the radial coordinate.

\section{Hydrostatic Equilibrium Equations}\label{Sec:III}
For all cases, the boundary condition of the metric potential $e^{-\lambda}$ is considered as the Schwarzschild exterior solution given as,
\begin{equation}
e^{-\lambda} = 1 - {2 \,m(r) \over r}\,,
\end{equation}
where $m(r)$ is the gravitational mass.

For the power-law form, using the Schwarzschild exterior solution with Eq.~ \eqref{eq:fRLM1Grr}--Eq.~\eqref{eq:fRLM1em}, the modified TOV-equations can be provided as
\begin{eqnarray}
\frac{dm}{dr} &=& 4\pi r^2 \Big[\rho+\sigma\, p (p + 2\rho) \Big], \label{eq3.2}\\[1ex]
\frac{dp}{dr} &=& -\frac{(p+\rho)\left[m + 4\pi r^3 p (1+\sigma\, p)\right]}{r^2(1 - 2m/r)}~. \label{eq3.3}
\end{eqnarray}
For the square-root exponential form, using the Schwarzschild exterior solution with Eq.~\eqref{eq:fRLM2Grr}--Eq.~\eqref{eq:fRLM2em}, the modified TOV equations are reduced to
\begin{eqnarray}
{dm\over dr} &=& 4 \pi  r^2 \rho + \left[2 \pi  \gamma  r^2 \rho   \sqrt{\beta \, p}-4 \pi  \gamma \, r^2 \rho  +2 \pi  \gamma \, r^2 p  \,\sqrt{\beta  \,p} \right]\,e^{-\sqrt{\beta \, p}}~~, \label{eq3.4}\\ [1ex]
{dp \over dr}  &=& -\frac{p+\rho}{r^2 (1-2 m/r)}\Big[\left(m+4 \pi  r^3 p\right)-4 \pi  \gamma \,  r^3 p\,e^{-\sqrt{\beta \, p}}\Big]~.\label{eq3.5}    
\end{eqnarray}
Using boundary condition and  Eq.~\eqref{eq:fRLTGtt}--Eq.~\eqref{eq:fRLTem}, the modified TOV equations of non-linear form of $f(\mathcal{R,L}_m,\mathcal{T})$ gravity become,
\begin{eqnarray}
\hspace{-1.5cm} {dm \over dr} &=& 4 \pi  r^2 \rho+ \frac{\alpha \, r^2}{4} \Big[2 p^2+ \rho(5 p -\rho) \Big]
\label{eq:fRLTtovm}, \label{eq3.6}\\ [1ex]
{dp \over dr} &=& -\frac{\bigl(p + \rho \bigr) \bigl(16 \pi + 5 \alpha \,p -  \alpha \,\rho \bigr) \big[4 m + p \,r^3 \bigl(16 \pi + 3 \alpha \,p -  \alpha \rho \bigr)\big]}{4  r^2 \big(1-2 m/r\big) \Bigl(16 \pi -  \alpha \big[\left({d\rho / dp}-6 \right) p + \rho \big]\Bigr)}~.
\label{eq3.7}
\end{eqnarray}
To calculate the modified TOV equations for each case, we must consider EoS of fully degenerate electron gas, which we will discuss in the next section.

\section{Equation of State}\label{Sec:IV}
To solve differential equation systems for each case such as, \eqref{eq3.2}--\eqref{eq3.3}, \eqref{eq3.4}--\eqref{eq3.5} and \eqref{eq3.6}--\eqref{eq3.7}, we need to consider an appropriate EoS that provides $p(\rho)$. For WDs, we considered Chandrasekhar EoS at zero temperature describing a completely degenerate relativistic gas as formulated by  \cite{Chandrasekhar_1931_456}. For a completely degenerate $e^-$ gas, the number of electrons $N$ in a finite volume $V$ in an arbitrary momentum $P$ and below the Fermi momentum $P_F$ are respectively given as
\begin{eqnarray}
{ N=V \times {8\pi \over h^3} \int_0^{P_F} P^2 dP=V \times {8\pi \over 3h^3}\,P_F^3 }~. \label{e4.1}
\end{eqnarray}
The pressure generated by the $e^-$ gas can be calculated as
\begin{eqnarray}
\hspace{-0.7cm} p = {1 \over 3V} \int_0^\infty N(P)\,P\,v_{_P}\,dP= {1 \over 3V} \int_0^{P_F} N(P)\,P\,v_{_P}\,dP = {1 \over 3V} \int_0^{P_F} N(P)\,P\,{\partial E \over \partial P}\,dP ,\label{e4.2}
\end{eqnarray}
where $v_{_P}$ and $E$ are the velocity and kinetic energy corresponding to the momentum $P$. For relativistic electrons, the kinetic energy can be written as
\begin{eqnarray}
E = m_ec^2\left[\left(1+{P^2 \over m_e^2c^2} \right)^{1/2} -1 \right] ~~~\mbox{or}~~~{\partial E \over \partial P}={P \over m_e}\left(1+{P^2 \over m_e^2c^2} \right)^{-1/2}~,
\end{eqnarray}
leading to 
\begin{eqnarray}
p={8\pi \over 3m_eh^3} \int_0^{P_F} {P^4dP \over \big(1+{P^2/m_e^2c^2} \big)^{1/2}}~,
\end{eqnarray}
which integrates by a variable transformation $\sinh \theta=P/m_ec$ as
\begin{eqnarray}
\hspace{-0.7cm} p={8\pi m_e^4c^5 \over 3h^3} \int_0^{\theta_F} \sinh^4 \theta\,d\theta={8\pi m_e^4c^5 \over 3h^3} \left[{\sinh^3\theta_F \cosh \theta_F \over 4}-{3 \sinh (2\theta_F) \over 16}+{3\theta_F \over 8} \right]~.
\end{eqnarray}
Using dimensionless Fermi momentum, $X_F:= P_F/m_e \,c$ where $m_e$ is the electron mass, the pressure can be written as
\begin{eqnarray}
p(X_F) &=& {\pi\, m_e^4 \,c^5 \over 3 \,h^3} \left[\left(2X_F^3 - 3X_F\right) \sqrt{X_F^2+1}\,+3\, \sinh^{-1}X_F\right]~.
\end{eqnarray}
The corresponding mass density is obtained from the $e^-$ number density, $n_e$ where
\begin{eqnarray}
    n_e={8\pi\,(m_e\,c)^3\over 3 \,h^3 } \,X_F^3~.
\end{eqnarray}
Since, $\rho=n_e \mu_e m_p$, the density can be written as
\begin{eqnarray}
\rho(X_F) ={8\pi\,\mu_e m_p(m_e\,c)^3\over 3 \,h^3 } \,X_F^3~~,
\end{eqnarray}
where  $m_e$ is the electron mass, $m_p$ the proton mass, $h$ is the usual Planck's constant and $\mu_e$ is the molecular weight per electron. For our article, we will consider C-O WDs, whose molecular weight, $\mu_e \approx 2$.


\begin{figure}
    \centering
    \includegraphics[width=\textwidth]{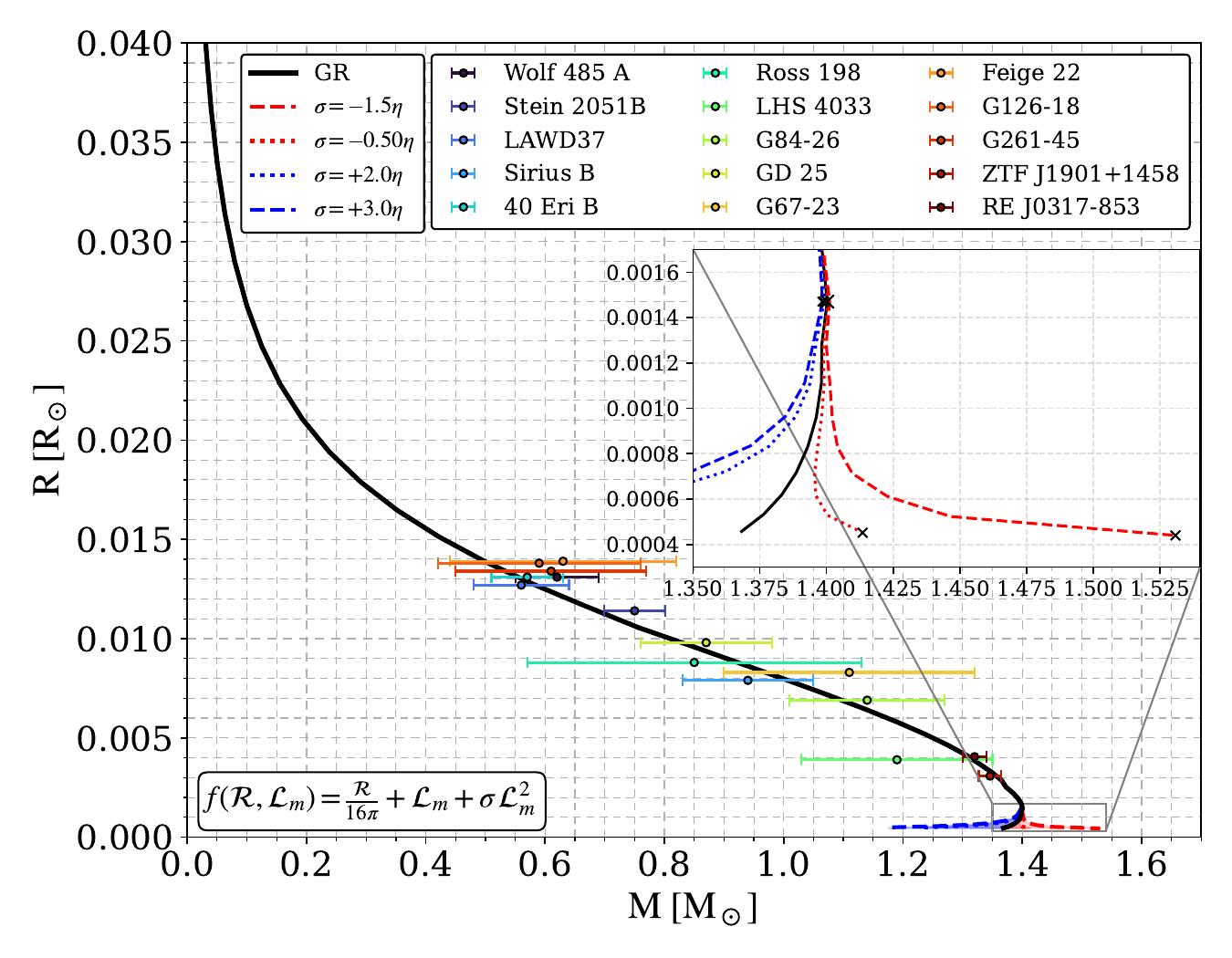}
    \caption{$\rm M-\rm R$ plot for WDs within the power-law $f(\mathcal{R},\mathcal{L}_m)$ gravity model. The solid black curve represents the predictions of GR, while the colored curves correspond to different values of the dimensionless coupling parameter $\sigma$, ranging from $-1.50\eta$ to $+3.0\eta$, with $\eta=10^{7}$.
 }
    \label{fig:MR_f(R,Lm)_M1}
\end{figure}


\begin{figure}[t]
    \centering
    \includegraphics[width=\textwidth]{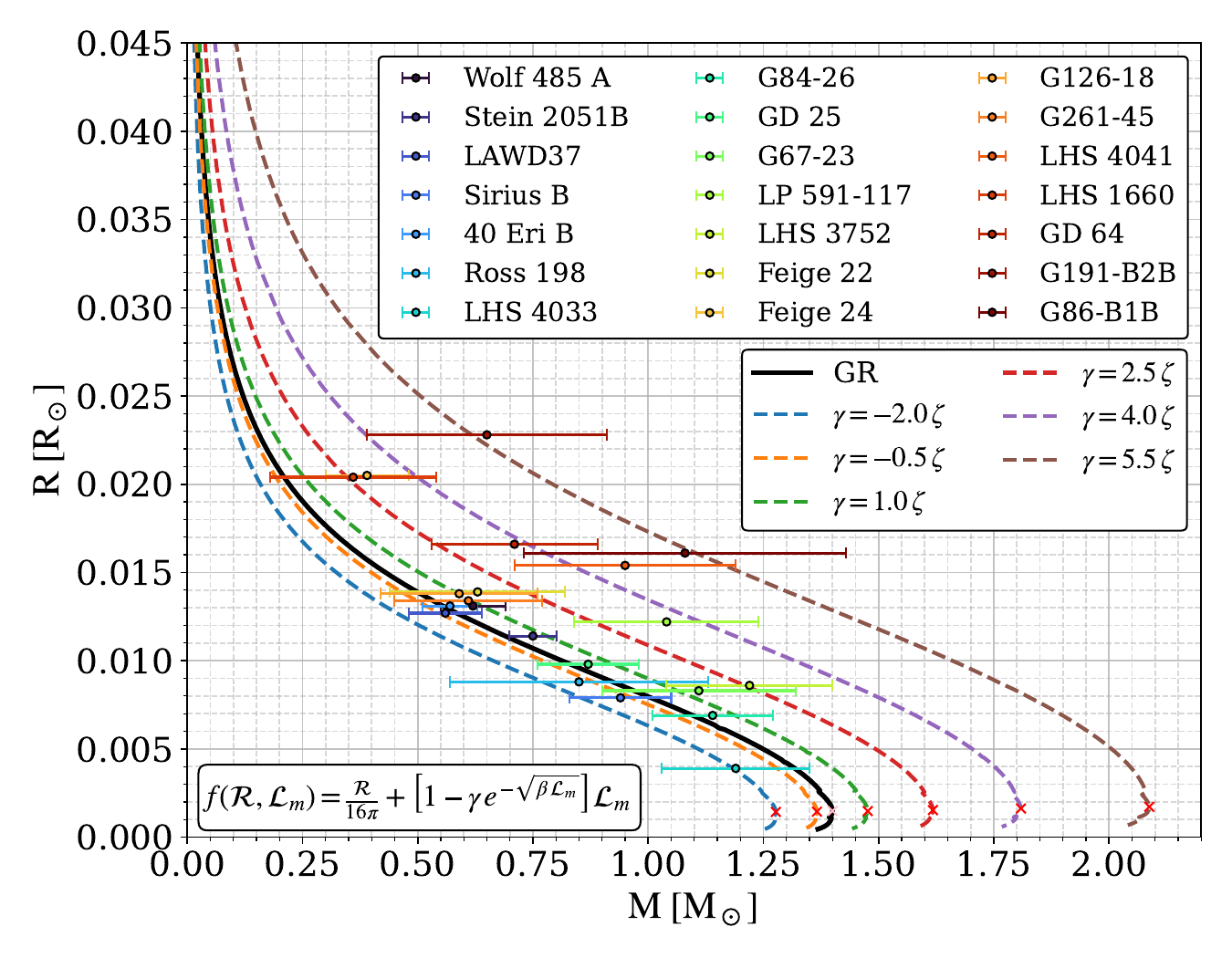}
    \caption{$\rm M-\rm R$ diagram for WDs in the $f(\mathcal{R},\mathcal{L}_m)$ gravity model with a square-root exponential coupling. The dimensional parameter $\beta$ has units of inverse pressure, $[\beta]=\mathrm{cm}^2\,\mathrm{dyne}^{-1}$, and is fixed to the dimensionless value $\tilde{\beta}=2$ throughout this work. The coupling parameter $\gamma$ varies from $-2.0\zeta$ to $5.5\zeta$, with $\zeta=10^{-1}$. The solid black curve represents the GR prediction, while the dashed curves correspond to different values of $\gamma$.
 }
    \label{fig:MR_f(R,Lm)_M2}
\end{figure}

\section{Mass-Radius Analysis}\label{Sec:V}
The WD configurations are computed using the fully degenerate EoS derived in the previous section.  The modified TOV equations \eqref{eq3.2}--\eqref{eq3.7} are solved numerically over the range of central densities, $\rho_c$, to obtain the corresponding mass-radius relations. { To numerically solve these equations,  we also utilize dimensionless physical variables as described in} Ref. \citep{Priyobarta_2025},

\begin{eqnarray}
 \tilde{r} = \frac{r}{r_g}, \quad \tilde{p} = \frac{p}{p_g}, \quad \tilde{\rho} = \frac{\rho}{\rho_g}, \quad \text{and} \quad \tilde{m} = \frac{m}{\rm M_\odot} ~,
    \label{eqn:scaling}
\end{eqnarray}
where the characteristic scales are defined as:
\begin{eqnarray}
r_g = \frac{G\, \rm M_\odot}{c^2}, \quad p_g = \frac{\rm M_\odot c^2}{r_g^3}, \quad \text{and} \quad \rho_g = \frac{\rm M_\odot}{r_g^3}~.
\end{eqnarray}
Here, $\rm M_\odot = 1.989 \times 10^{33}\, \text{g}$ represents the mass of the Sun, $c = 2.997 \times 10^{10}\, \text{cm/s}$ indicates the speed of light, and $G = 6.674 \times 10^{-8}\, \text{dyne} \, \text{cm}^2 / \text{g}^2$ is the gravitational constant. The coupling parameters are likewise transformed into dimensionless quantities according to
\begin{equation}
\bm{\tilde{\alpha} = \alpha\, p_g,}\qquad
\tilde{\beta} = \beta\, p_g,\qquad
\tilde{\sigma} = \sigma\, p_g.
\end{equation}
Since $\gamma$ is dimensionless, it remains unchanged.
For notational simplicity, the tildes are omitted hereafter, and all
quantities are understood to be dimensionless unless otherwise stated. After we convert in the dimensionless parameter and apply the boundary conditions at $r=0$ as
\begin{eqnarray}
\tilde{\rho}\,(0) = \rho_c \quad \text{and} \quad \tilde{m}\,(0) = 0 ~,
\end{eqnarray}
 where $\rho_c$ is the central density.  We compute the integration using the Runge-Kutta 45 (RK45) method available in the `scipy.integrate' package in Python until the pressure decreases to zero at the point where $p(\rm R) = 0$. We consider this point to be the surface of the WDs and denote its radius as $\rm R$.  Finally, using this radius $\rm R$, we calculate the maximum mass attained along the WDs sequence. The adaptive step size is automatically controlled by the error estimation procedure of the Runge--Kutta integrator using the default tolerances of the SciPy implementation.  By repeatedly calculating the structure of WDs over the range of central density $1\times 10^{7}-2.4 \times 10^{12}\,\mathrm{g}/ \mathrm{cm}^3$, we obtain $\rm M-\rm R$ relation for each of the model of both $f(\mathcal{R,L}_m)$ and $f(\mathcal{R,L}_m, \mathcal{T})$ gravity.  We also compare our theoretical predictions with observational constraints for several well-known WDs taken from Ref.~\citep{Bedard_2017}. These observational data are shown for qualitative comparison only and are not used to constrain the model parameters. For models predicting an extended mass range beyond the standard Chandrasekhar limit, additional massive WD candidates such as LHS 3752, G86$-$B1B, ZTF J1901+1458 and RE J0317$-$853 are included to assess the compatibility of the theoretical $\rm M-\rm R$ relations with the most massive observed WDs.

\begin{figure}[h]
\centering
\includegraphics[width=\textwidth]{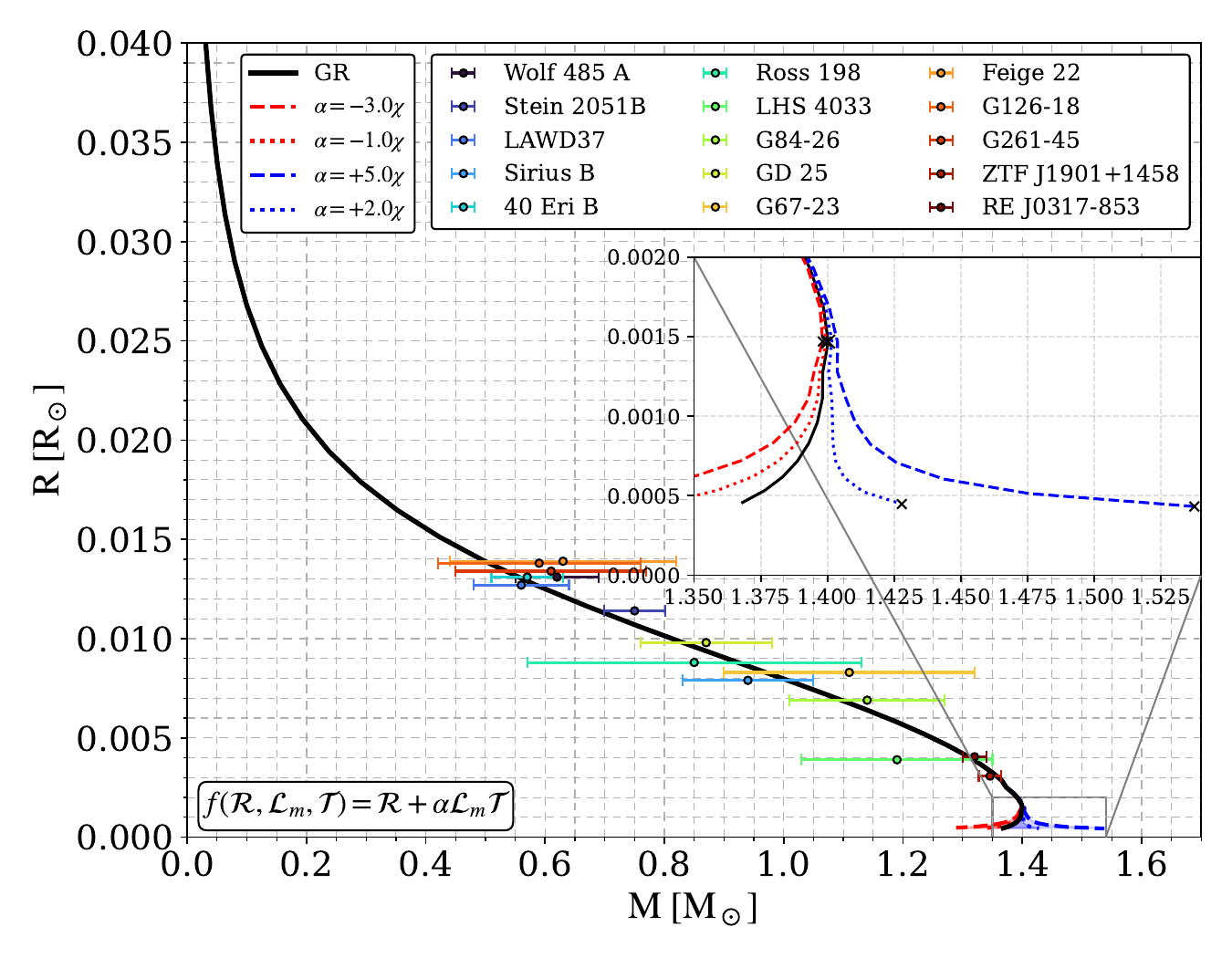}
\caption{$\rm M-\rm R$ plot for WDs within the non-linear $f(\mathcal{R},\mathcal{L}_m,\mathcal{T})$ gravity model. The solid black curve represents the GR prediction, whereas the colored curves correspond to the dimensionless coupling parameter values $\alpha$ ranging from $-3\chi$ to $+5\chi$  with $\chi=10^{6}$.} 
    \label{fig:MR_f(R,Lm,T)_M1}
\end{figure}

In the power-law model of $f(\mathcal{R,L}_m)$ gravity, we examined values of $\sigma$ ranging from $-1.5\eta$ to $+3.0\eta$, as illustrated in Figure~\ref{fig:MR_f(R,Lm)_M1}. The effects of the $\sigma \mathcal{L}_m^2$ term are particularly noticeable at high densities, specifically for densities $\rho_c \geq 10^{10} \mathrm{g/cm^3}$. Notably, with negative values of $\sigma$, it is possible to achieve a higher mass compared to the GR and/or Newtonian scenario. For $\sigma = -1.5\eta$ and $\sigma = -0.50\eta$, the model can attain maximum masses along the corresponding equilibrium sequences up to $1.52\mathrm{M}_\odot$ and $1.41\mathrm{M}_\odot$, respectively. Further, we also provided the maximum mass and radii in units of solar mass and solar radii for the different values of $\sigma$ in Table \ref{tab1:fRLm_M1}.

{In the case of the square-root exponential model of $f(\mathcal{R,L}_m)$ gravity, the parameter $\beta$ is fixed at $\beta=2$, while the parameter $\gamma$ is varied, as shown in the Figure~\ref{fig:MR_f(R,Lm)_M2}.} We have considered both positive and negative values, ranging from $-2.0\zeta$ to $5.5\zeta$, along with GR. The effects of the square-root exponential terms are clearly evident in the model, where positive values of $\gamma$ can result in higher mass. With the values of $\gamma= 5.5\zeta$, the maximum mass attained along the corresponding sequence reaches up to $2.08\mathrm {M}_\odot$, which is higher than the maximum masses obtained for the GR and/or Newtonian sequences. We also provided the maximum mass and radii of each value of $\gamma$ in Table \ref{tab1:fRLm_M2}.

Finally, for the non-linear model of  $f(\mathcal{R}, \mathcal{L}_m, \mathcal{T})$ gravity, we have plotted $\rm M-\rm R$ curve, as shown in Figure~ \ref{fig:MR_f(R,Lm,T)_M1}. By varying the values of $\alpha$ with the ranges from $-3.0\chi$ to $+5.0 \chi$, we examined how the  $\rm M-\rm R$ curve changes and the effects of the coupling term is particularly noticeable at high densities, specifically for densities $\rho_c \geq 10^{10} \mathrm{g/cm^3}$. as similar to the power-law model of $f(\mathcal{R,L}_m)$ gravity. { \cite{Otoniel_2026} also provided similar plots in the non-linear form of $f(\mathcal{R,L}_m, \mathcal{T})$ gravity which also shown maximum mass can be generated with the positive values of $\alpha$.} With the value of $\alpha=+5.0\,\chi$, one can generate the maximum mass of $1.53\mathrm{M}_\odot$. Table \ref{tab1:fRLmT}  provides the maximum mass and corresponding radii for each value of the parameter. 

In all the models examined, it is evident that the modified gravity corrections can significantly increase the maximum mass attained along the corresponding WD sequences relative to GR and/or Newtonian gravity. {We have also include the observational data for WDs to provide a crucial benchmark for evaluating the theoretical $\rm M$--$\rm R$ relationships. The theoretical $\rm M$--$\rm R$ relations are presented alongside the observational data for qualitative comparison. These modified gravity models align well with a wide range of observed WD masses and radii, especially for those higher masses, like ZTF J1901+1458 and RE J0317-853.  Moreover, the $\mathrm{M}-\mathrm{R}$ relation predicted by the square-root exponential model of $f(\mathcal{R},\mathcal{L}_m)$ gravity appears to be visually closer to the observed properties of certain WDs, including G86-B1B and GD 64, than those predicted by the non-linear and power-law models of $f(\mathcal{R},\mathcal{L}_m,\mathcal{T})$ and $f(\mathcal{R},\mathcal{L}_m)$ gravity, respectively.}

\section{Stability Analysis}\label{Sec:VI}
To develop effective models, we examine different stability criteria to ensure that the systems do not collapse. Stability can be evaluated through various approaches. Here, we have three approaches such as (i) static stability criterion, (ii) compactness and gravitational redshift, and (iii) adiabatic index.

\subsection{Static Stability Criterion}
We analyses the stability using the general relativistic case,  $ d\rm M / d \rho_c >0$, i.e., as the central density increases, the mass increases until it reaches the maximum mass. For the power-law model of $f(\mathcal{R,L}_m)$ gravity, gravitational mass also increases monotonically  with the central density until the maximum mass configuration is reached. Over the ranges of $\sigma$  from $-1.5 \eta $ to  $ +3.0\eta$, we have plotted the $\rm M-\rho_c$ curve as shown in Figure~ \ref{fig:MRho_f(R,Lm)_M1}. For all the consideration of $\sigma$, the given conditions hold {up to} the maximum mass configuration. Positive values of $\sigma$, the central density of the maximum mass hold $\rho_c\approx 10^{10} \, \mathrm{g/cm^3} $  whereas negative values hold up to $\rho_c\approx 10^{12}\, \mathrm{g/cm^3}$.

Using the same criterion, we also analyze the square-root exponential form of $f(\mathcal{R,L}_m)$ gravity. The $\rm M-\rho_c$  plot for GR and for the coupling parameter $\gamma$ in the range of $-2.0\zeta$ to $+5.5\zeta$ are shown in Figure~\ref{fig:MRho_f(R,Lm)_M1}. For both GR and all considered values of  $\gamma$, gravitational mass also increases monotonically with central density until it reaches maximum mass, which satisfies the condition $d \rm M/d\rho_c >0$. For all cases, the central density of the maximum mass lies in the range of $10^{10}<\rho_c<10^{11} \,\, \mathrm{g/cm^3}$. Finally, in the non-linear model of $f(\mathcal{R,L}_m,\mathcal{T})$ gravity, the gravitational mass monotonically increases with central density until it reaches its maximum mass. We have plotted the $\rm M-\rho_c$ curve for the range of $\alpha$ from $-3.0 \chi$ to $+5.0 \chi$, as illustrated in Figure~\ref{fig:MRho_f(R,Lm,T)_M1}. For every value of $\alpha$ considered, the specified conditions remain valid up to the maximum mass configuration. For negative values of $\alpha$, the central density at maximum mass is approximately $\rho_c \approx 10^{10} \, \mathrm{g/cm^3}$, while for positive values, it reaches about $\rho_c \approx 10^{12} \, \mathrm{g/cm^3}$. For all the models considered, the coupling parameter satisfies the stability criterion, i.e., $d \rm M/d\rho_c>0$. Central density as high as $\sim 10^{12}\,\, \mathrm{g/cm^3}$  are obtained in the case of power-law model of $f(\mathcal{R,L}_m)$  gravity and non-linear model of $f(\mathcal{R,L}_m,\mathcal{T})$ gravity. In contrast, the square-root exponential model of $f(\mathcal{R,L}_m)$ gravity  yields lower central densities in the ranges $10^{10}-10^{11}\,\, \mathrm{g/cm^3}$.

\begin{figure}
\centering
\includegraphics[width=8cm, height=7cm]{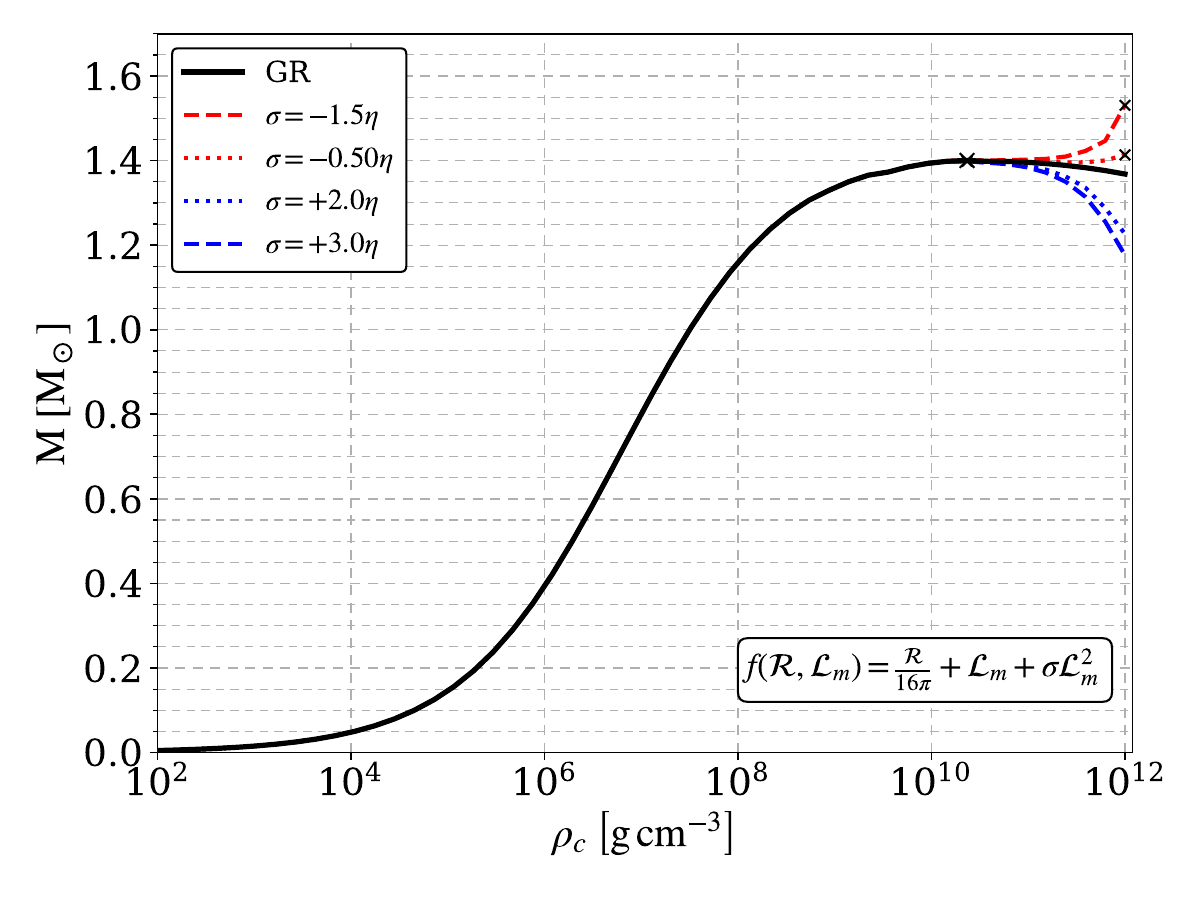}~~~\includegraphics[width=8cm, height=7cm]{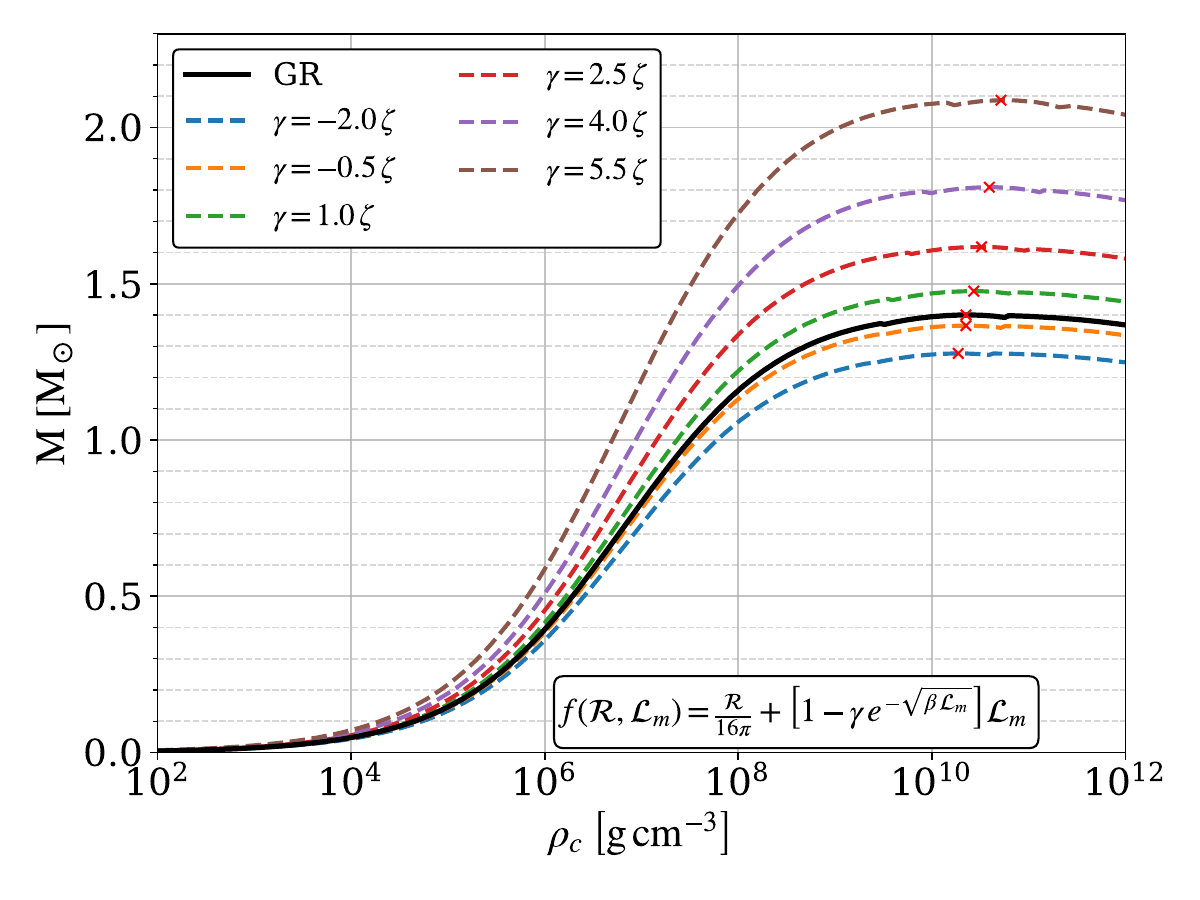}
\caption{$\rm M-\rho_c$ plots within the framework of the power-law and square-root exponential models of $f(\mathcal{R,L}_m)$ gravity. The solid black line represents the GR, while colored curves represent the different values of $\sigma$  and $\gamma$.}
\label{fig:MRho_f(R,Lm)_M1}
\end{figure}

\begin{figure}
    \centering
    \includegraphics[width=10cm, height=8cm]{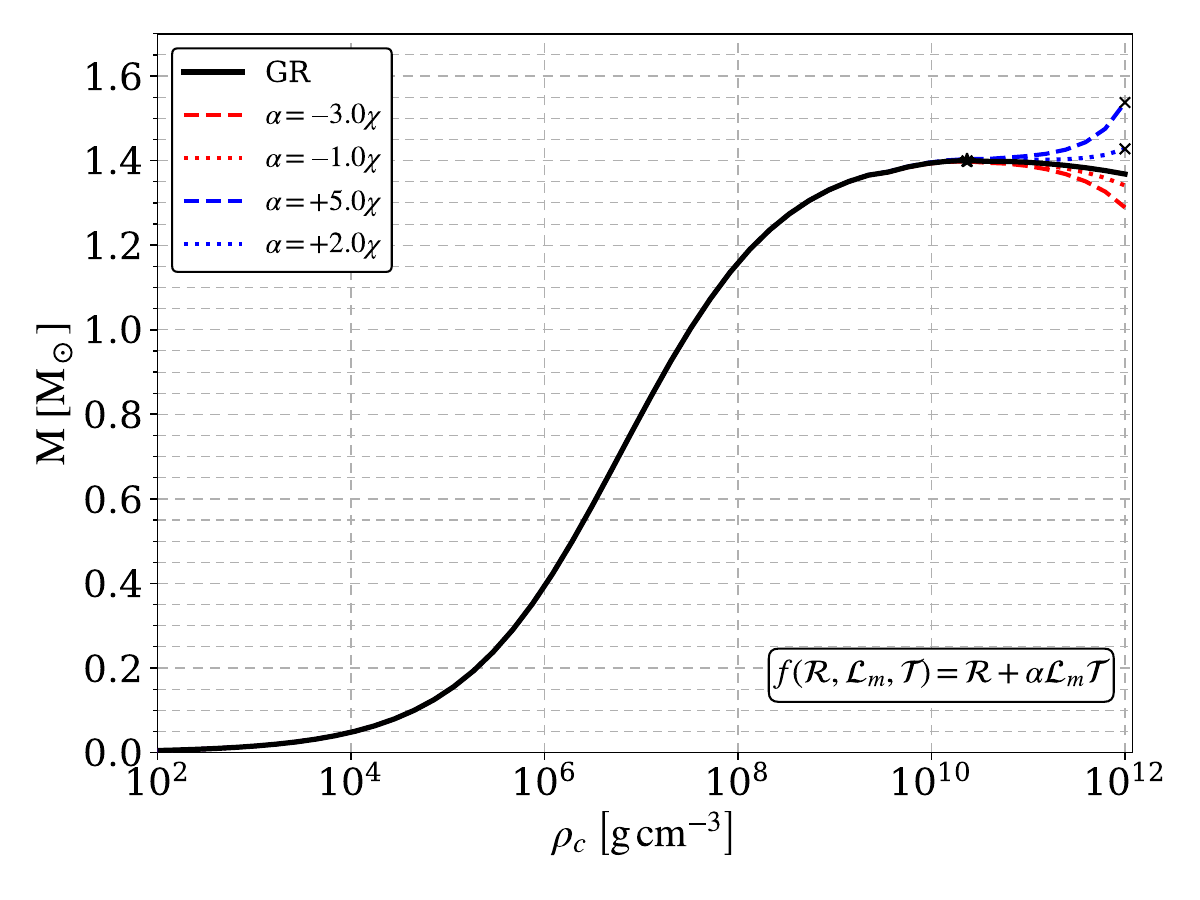}
    \caption{$\rm M-\rho_c$ plots within the non-linear model of $f(\mathcal{R,L}_m,\mathcal{T})$ gravity. The solid black lines represent the GR, whereas the colored curves represent the various values of the coupling parameter $\alpha$.}
    \label{fig:MRho_f(R,Lm,T)_M1}
\end{figure}

\begin{figure}
\includegraphics[width=8cm, height=7cm]{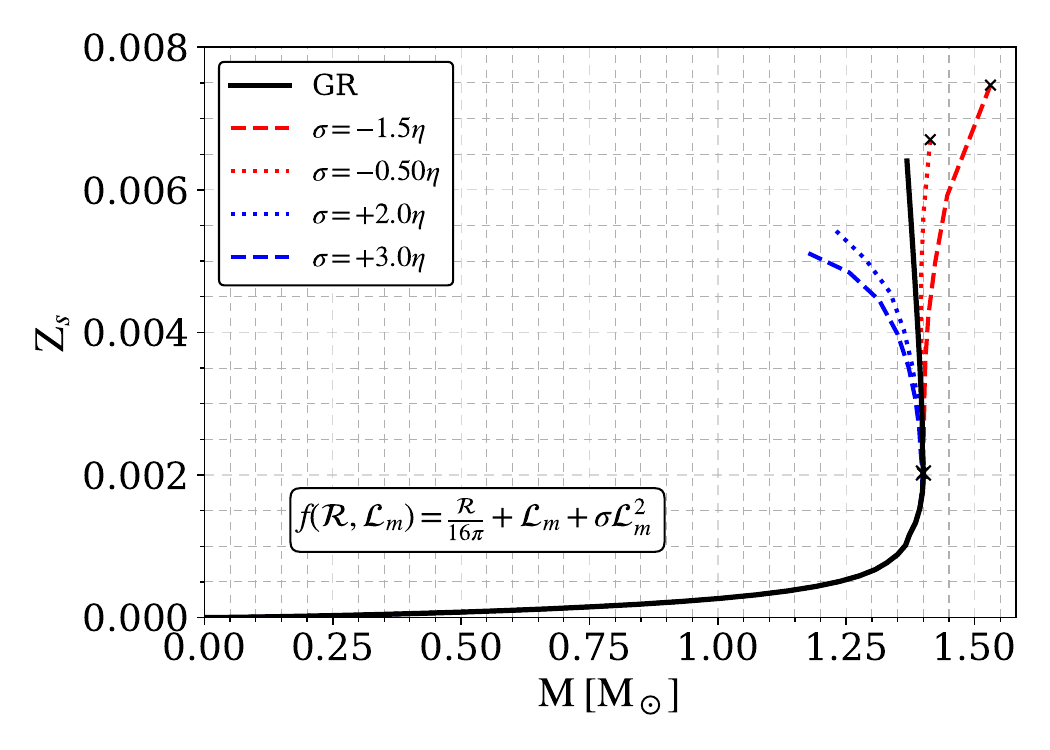}~~~\includegraphics[width=8cm, height=7cm]{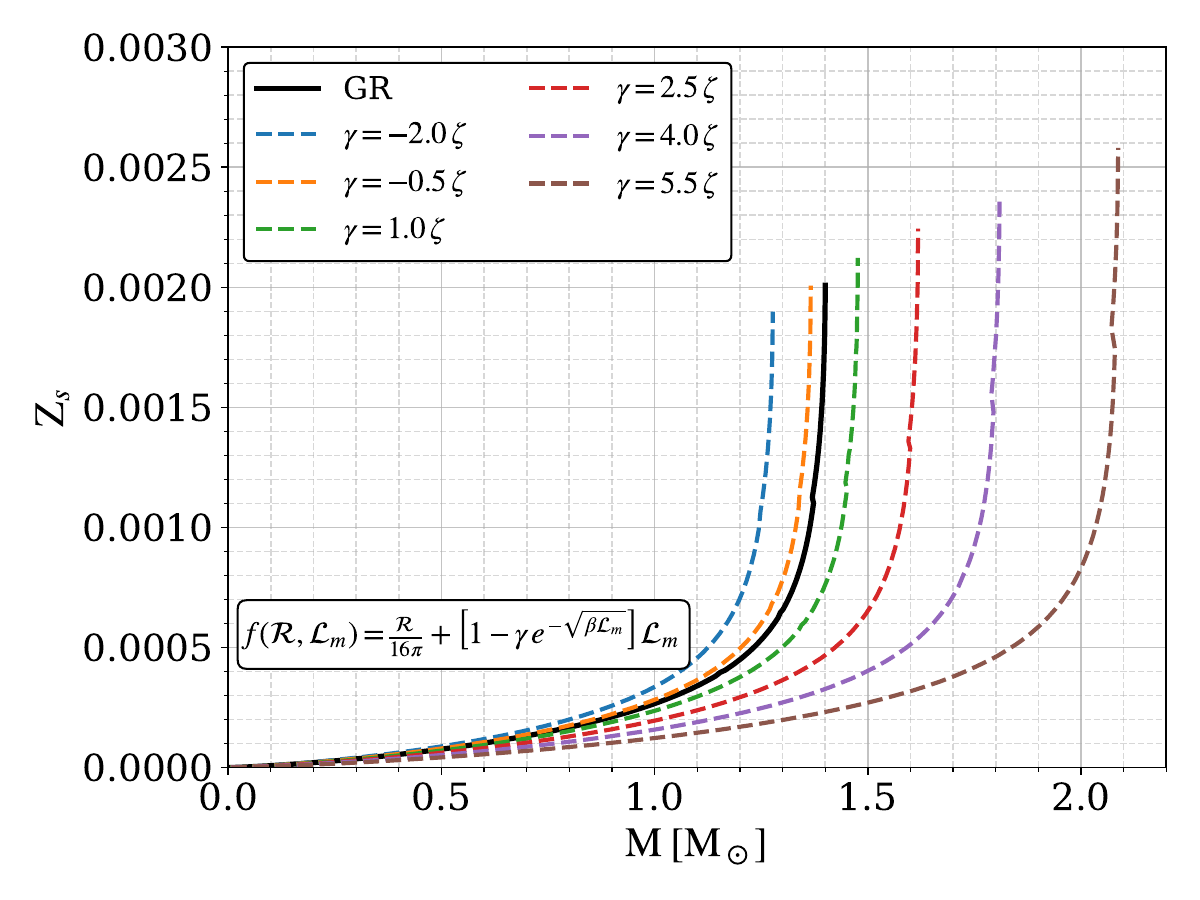}
\caption{$\rm Z_s-\rm M$ plots within the framework of the power-law and square-root exponential model models of $f(\mathcal{R,L}_m)$ gravity. The solid black lines denote GR, while the colored curves illustrate different values of the coupling parameter $\sigma$ and $\gamma$.}\label{fig:ZM_f(R,Lm)_M1}
\end{figure}

\begin{figure}
    \centering
    \includegraphics[width=11cm, height=8cm]{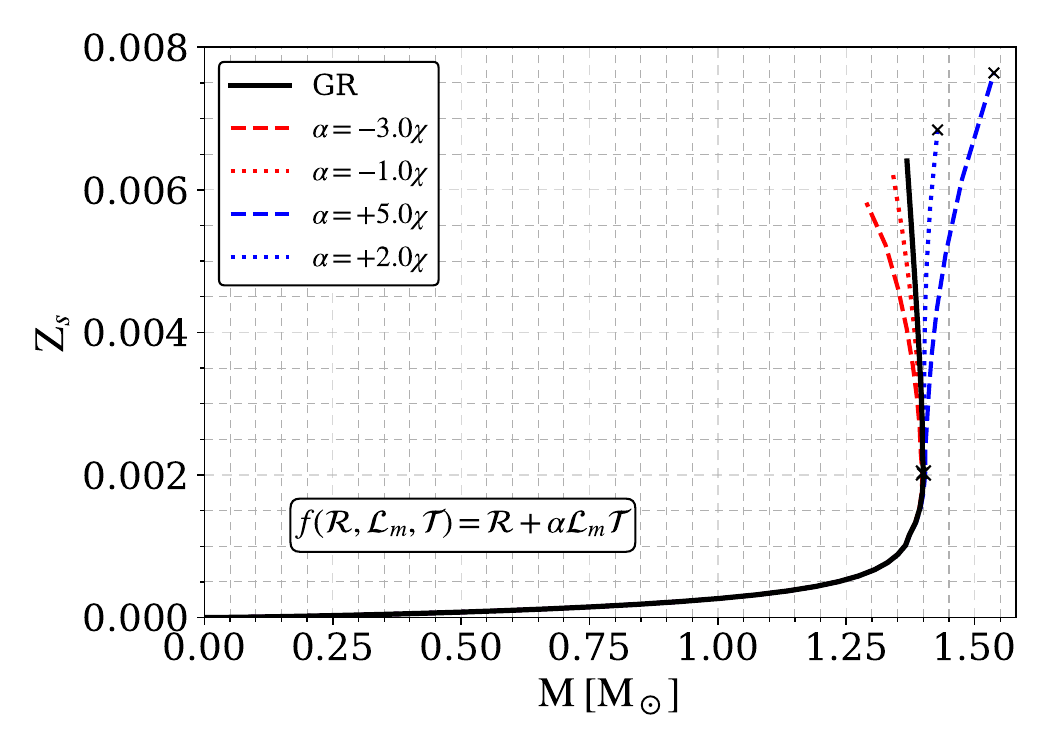}
    \caption{$\rm Z_s-\rm M$ plot for WDs within the non-linear model of  $f(\mathcal{R}, \mathcal{L}_m, \mathcal{T})$ gravity. The solid black curve represents the predictions of GR, while the colored curves depict various values of the coupling parameter  $\alpha$.}
    \label{fig:ZM_f(R,Lm,T)_M1}
\end{figure}

It is important to highlight that the stability criterion $d\rm M/d\rho_c > 0$ mentioned earlier is derived from a series of equilibrium states and indicates stability against radial disturbances up to the maximum mass point. However, for the extremely high central densities found in the gravity models we consider ($\rho_c \sim 10^{12} \, \mathrm{g/cm^3}$), additional microphysical effects not included in the present Chandrasekhar EoS may become relevant \citep{Shapiro_1983,Rotondo_2011,Salpeter_1969}. Specifically, processes such as inverse beta decay, neutronization, and pycnonuclear fusion could alter chemical composition and reduce the electron degeneracy pressure that supports WDS \citep{Boshkayev_2012}. Thus, while the configurations meet the equilibrium stability criterion ($d\rm M/d\rho_c > 0$), they should be viewed as theoretical solutions within the confines of the chosen EoS. A comprehensive evaluation of the physical plausibility of these ultra high density configurations would require consideration of microphysical effects, which are not addressed in the current study.

\subsection{Compactness and Gravitational Redshift}

The gravitational redshift enables us to measure the strength of the gravitational field around WDs, thereby deepening our understanding of gravitational behavior in extreme environments. Additionally, we examine the correlation between surface gravitational redshift $\rm Z_s$ and mass for WDs. The surface gravitational redshift is defined as
\begin{eqnarray}
    \rm Z_s = {1 \over \sqrt{1-2\, \mathrm{C}}}-1 ~.
\end{eqnarray}
Here, $\mathrm{C}= \rm M/\rm R$  is the compactness of the stellar structures. Due to the relatively large radii and lower masses compared to other compact objects like NS, WDs have very small compactness values, leading to low surface gravitational redshifts on the order of $\mathcal{O}(10^{-3})$.

We have analyzed $\rm Z_s-\rm M$ for the power-law model of the $f(\mathcal{R,L}_m)$ gravity, considering both GR and parameter $\sigma$ within the same ranges as previously adopted. The corresponding results are shown in the Figure~\ref{fig:ZM_f(R,Lm)_M1}. As expected, the very small values of compactness and redshift remained small i.e., $\mathcal{O}(10^{-3})$. Higher redshift values are obtained in the case of negative values of $\sigma$. The numerical values of compactness and redshift are also provided in Table \ref{tab1:fRLm_M1}. For the square-root exponential model of $f(\mathcal{R,L}_m)$  gravity, we have plotted the $\rm Z_s-\rm M$ relation for both GR and for parameter $\gamma$  within the same ranges as previously considered as shown in the Figure~\ref{fig:ZM_f(R,Lm)_M1}. In this case, positive values of $\gamma$ provided a higher redshift as compared to the GR case. We have also provided the compactness and redshift values in Table \ref{tab1:fRLm_M2}. Finally, we analyzed the $\rm Z_s-\rm M$ for the GR and for the coupling parameter $\alpha$ of the non-linear model of $f(\mathcal{R,L}_m,\mathcal{T})$ gravity within the same ranges. In contrast with the power-law model of $f(\mathcal{R,L}_m)$ gravity, in this model, positive values of $\alpha$ provided the higher redshift as shown in the Figure~\ref{fig:ZM_f(R,Lm,T)_M1}. The numerical values of compactness and redshift are also provided in the Table \ref{tab1:fRLmT}.

For all the models considered, the compactness and redshift are significantly smaller than those of the NS. Due to their smaller magnitudes, the numerical values of compactness and redshift are nearly identical. Relatively higher values are obtained in power-law model of $f(\mathcal{R,L}_m)$ gravity and in non-linear model of $f(\mathcal{R,L}_m,\mathcal{T})$ gravity when compared to the square-root exponential model of $f(\mathcal{R,L}_m)$ gravity.

\subsection{Adiabatic Index}

We emphasize an important aspect of stability analysis i.e., the adiabatic index, $\Gamma(r)$. This parameter is crucial in thermodynamics, particularly under adiabatic conditions, where it defines the relationship between pressure and density. The adiabatic index illustrates how pressure changes with density during these processes. Using the variational method, \cite{Chandrasekhar_1964} showed that for dynamical stability to be achieved, the condition $\Gamma > 4/3$ must be satisfied. In the relativistic regime, general relativistic corrections further enhance the onset of instability, necessitating that $\Gamma$ exceeds the Newtonian limit. The relativistic formulation of the adiabatic index and its significance in stellar stability were later thoroughly discussed by  \cite{Herrera_1989} and   \cite{Misner_1973}.
\begin{eqnarray}
\Gamma =\left( {\rho + p \over p}\right)~{dp\over d \rho}~,
\end{eqnarray}
We have analyzed $\Gamma-r$ for the power-law model of the $f(\mathcal{R,L}_m)$ gravity, considering both GR and parameter $\sigma$ within the same ranges as previously adopted. The corresponding results are shown in the Figure~\ref{fig:Adiabatic_f(R,Lm)_M1}. { For both the GR and all considered values of  $\sigma$, one can clearly observe that the adiabatic index $\Gamma>4/3$, throughout the stellar interior of WDs, indicating that the configurations satisfy the adiabatic stability criterion considered in the present work.} For the square-root exponential model of $f(\mathcal{R,L}_m)$  gravity, we have plotted the $\Gamma-r$ relation for both GR and for parameter $\gamma$  within the same ranges as previously considered, as shown in the {Figure~ \ref{fig:Adiabatic_f(R,Lm)_M1}. Here, the dotted line represents the value of $\Gamma=4/3$, and all the curves lie above this limit, satisfying the adiabatic stability criterion.} Finally, we analyzed the $\Gamma-r$ for the GR and for the coupling parameter $\alpha$ of the non-linear model of $f(\mathcal{R,L}_m,\mathcal{T})$ gravity within the same ranges (Figure~\ref{fig:Adiabatic_f(R,Lm,T)_M1}). The horizontal dashed line represents the critical threshold $\Gamma=4/3$. {It is evident that, for GR and for all considered values of $\alpha$, the adiabatic index remains above this limit throughout the stellar interior, indicating that the configurations satisfy the adiabatic stability criterion.} For all the models considered, relatively higher values are obtained in power-law model of $f(\mathcal{R,L}_m)$ gravity and in non-linear model of $f(\mathcal{R,L}_m,\mathcal{T})$ gravity when compared to the square-root exponential model of $f(\mathcal{R,L}_m)$ gravity. The dynamical stability of the WDs is examined exclusively through the adiabatic criterion for the considered stellar configuration. By analyzing the adiabatic index $\Gamma$ as a function of the radial coordinate, we plot the $\Gamma-r$ profile for both GR and for parameter ranges of the considered models. In all the cases, the adiabatic index remains above the critical value, i.e., $\Gamma>4/3$ throughout the interior of the stellar structure, suggesting that the configurations satisfy the adiabatic stability condition.

\begin{figure}
\includegraphics[width=8cm, height=7cm]{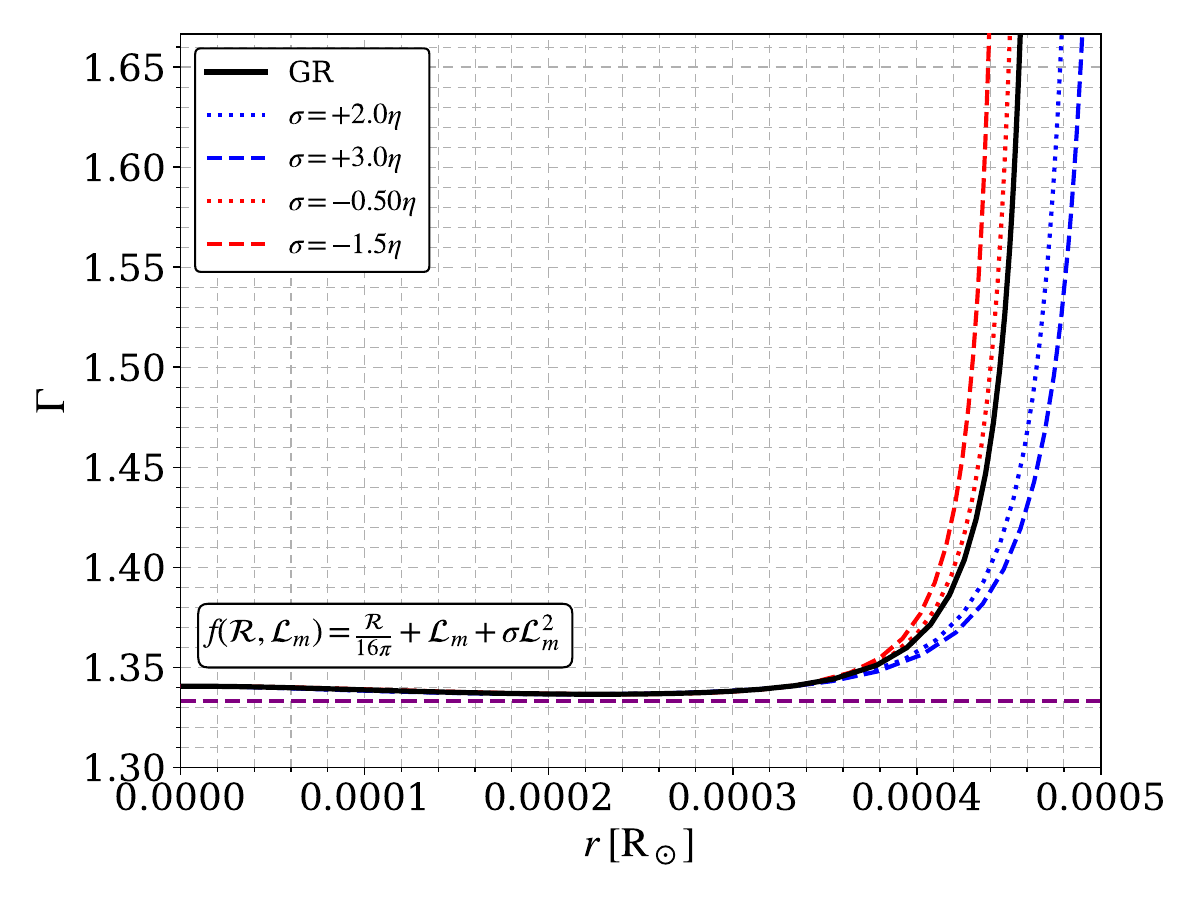}
\includegraphics[width=8cm, height=7cm]{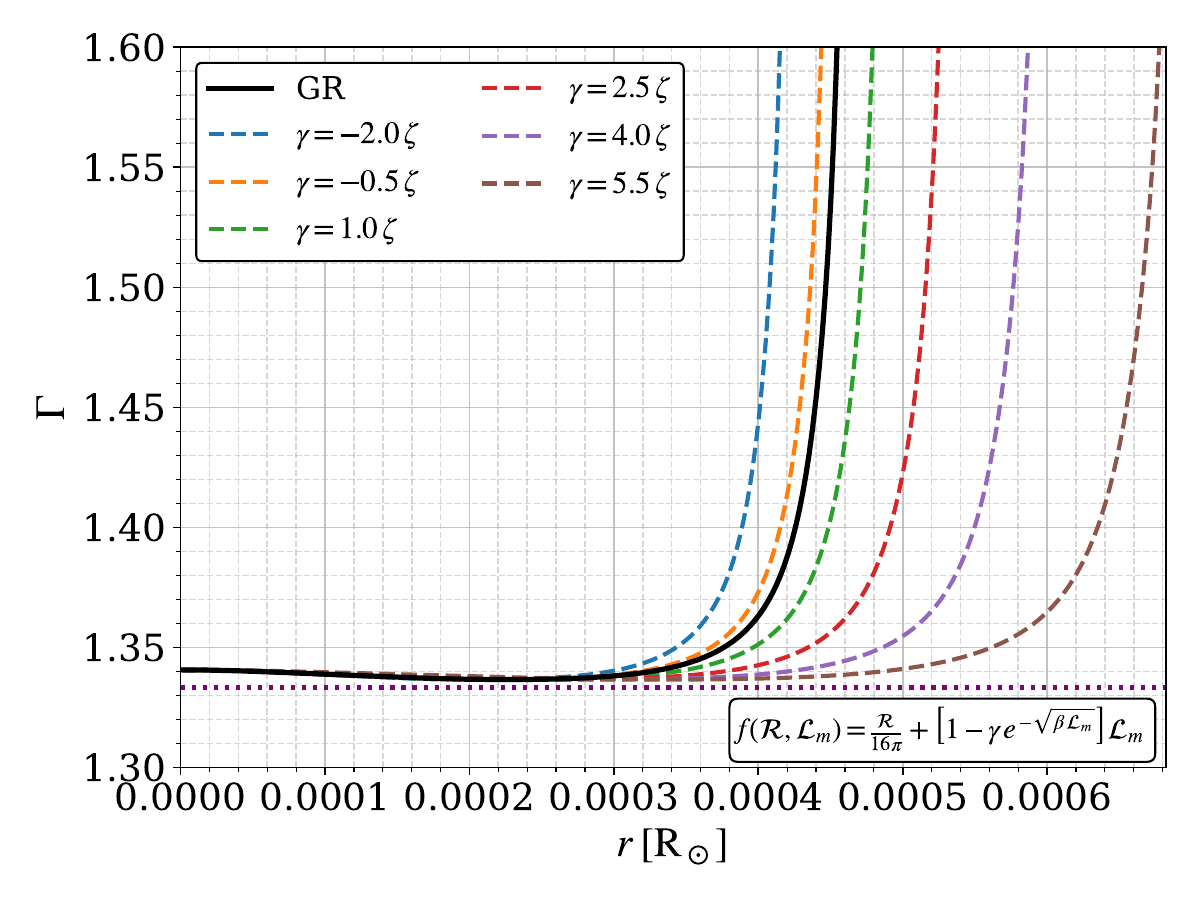}
\caption{Adiabatic Index, $\Gamma$ as a function of radial coordinate within power-law and square-root exponential models of $f(\mathcal{R,L}_m)$ gravity. The solid black line denotes the GR, while the colored lines represent the different values of $\sigma$ and $\gamma$. Here, the purple dashed line represents $\Gamma=4/3$ below which a stellar object will be unstable.} \label{fig:Adiabatic_f(R,Lm)_M1}
\end{figure}

\begin{figure}
    \centering
    \includegraphics[width=11cm, height=8cm]{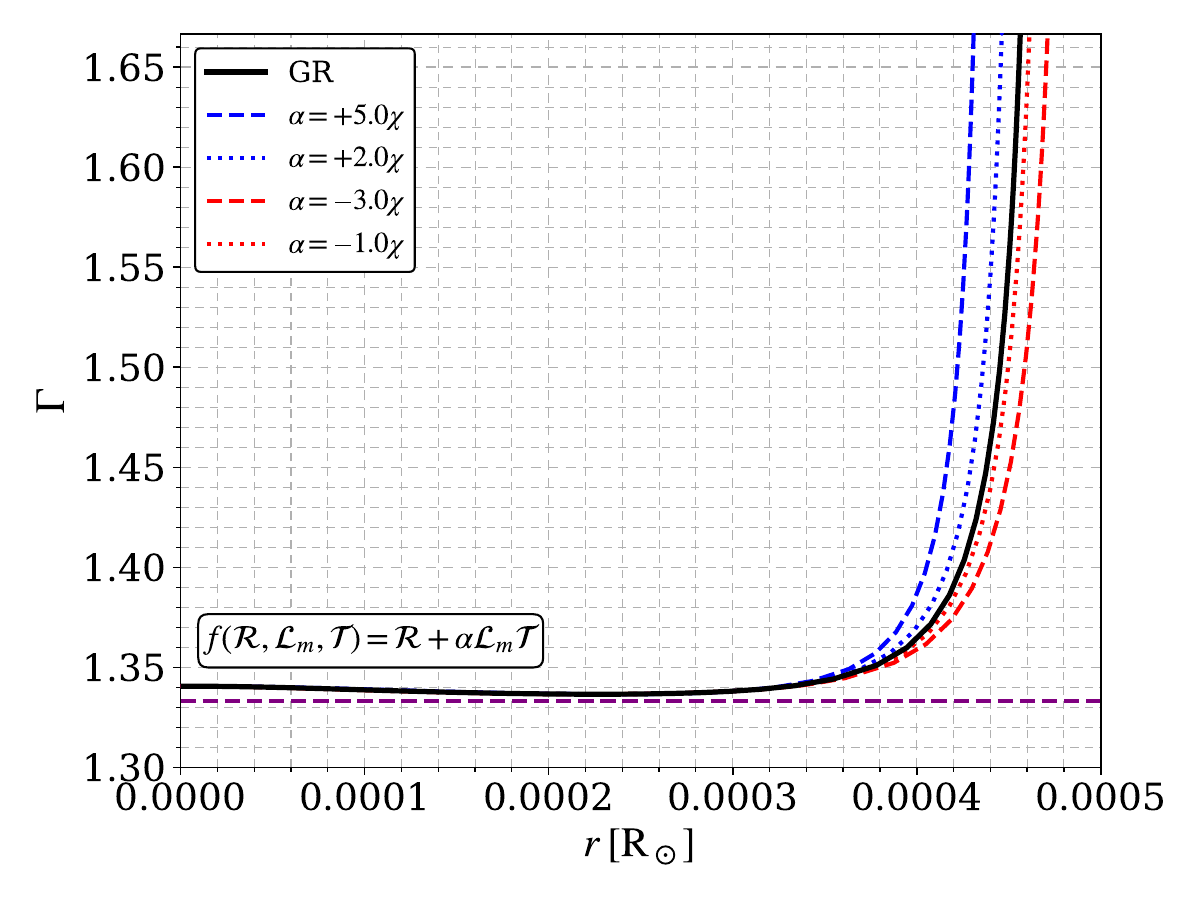}
    \caption{Adiabatic Index, $\Gamma$ as a function of radial coordinate within non-linear model of $f(\mathcal{R,L}_m,\mathcal{T})$ gravity. The solid black line denotes the GR, while the colored lines represent the different values of $\alpha$. Here, the purple dashed line represents $\Gamma=4/3$ below which a stellar object will be unstable.}
    \label{fig:Adiabatic_f(R,Lm,T)_M1}
\end{figure}

\begin{table}[htp!]
\centering
\caption{Structure properties of WDs configurations in $f(\mathcal{R,L}_m)$ gravity: Power-law model}
\label{tab1:fRLm_M1}
\renewcommand{\arraystretch}{1.2}
\begin{tabular}{|c|c|c|c|c|}
\hline
Parameter 
& $\mathrm{M}_{\text{max}}$  $(\mathrm{M}_\odot)$ 
& $\mathrm{R}_{\text{max}}$ $(\mathrm{R}_\odot \times 10^{-3})$ 
& $C\times10^{-3}$
&  $\rm Z_s \times10^{-3}$\\
\hline
$\sigma=-1.50\eta$ & 1.52 & 0.439 & 7.385 & 7.468 \\
\hline
$\sigma=-0.50\eta$ & 1.41 & 0.452 & 6.636 & 6.703 \\
\hline
$\sigma=0$ (GR)    & 1.40 & 1.47 & 2.021 & 2.027 \\
\hline
$\sigma=+2.0\eta$ & 1.39 & 1.47 & 2.019 & 2.025 \\
\hline
$\sigma=+3.0\eta$ & 1.38 & 1.47 & 2.018 & 2.024 \\
\hline
\end{tabular}
\end{table}
\begin{table}[htp!]
\centering
\caption{Structure properties of WDs configurations in $f(\mathcal{R,L}_m)$ gravity: Square-root exponential model.}
\label{tab1:fRLm_M2}
\renewcommand{\arraystretch}{1.2}
\begin{tabular}{|c|c|c|c|c|}
\hline
Parameter 
& $\mathrm{M}_{\text{max}}$  $(\mathrm{M}_\odot)$ 
& $\mathrm{R}_{\text{max}}$ $(\mathrm{R}_\odot \times 10^{-3})$ 
& $C\times10^{-3}$
&  $\rm Z_s \times10^{-3}$\\
\hline
$\gamma=-2.0\zeta$ & 1.27 & 1.43 & 1.893 & 1.898 \\
\hline
$\gamma=-5.0\zeta$ & 1.36 & 1.45 & 2.001 & 2.007 \\
\hline
$\gamma=0$ (GR)         & 1.40 & 1.47 & 2.021 & 2.027 \\
\hline
$\gamma=+1.0\zeta$ & 1.47 & 1.48 & 2.116 & 2.122 \\
\hline
$\gamma=+2.5\zeta$ & 1.61 & 1.53 & 2.237 & 2.245 \\
\hline
$\gamma=+4.0\zeta$ & 1.80 & 1.62 & 2.365 & 2.374 \\
\hline
$\gamma=+5.5\zeta$ & 2.08 & 1.72 & 2.570 & 2.580 \\
\hline
\end{tabular}
\end{table}
\begin{table}[htp!]
\centering
\caption{Structure properties of WDs configurations in $f(\mathcal{R,L}_m,\mathcal{T})$ gravity: Non-linear model.}
\label{tab1:fRLmT}
\renewcommand{\arraystretch}{1.2}
\begin{tabular}{|c|c|c|c|c|}
\hline
Parameter 
& $\mathrm{M}_{\text{max}}$  $(\mathrm{M}_\odot)$ 
& $\mathrm{R}_{\text{max}}$ $(\mathrm{R}_\odot \times 10^{-3})$ 
& $C\times10^{-3}$
&  $Z_s \times10^{-3}$\\
\hline
$\alpha=-3.0\chi$ & 1.398 & 1.471 & 2.017 & 2.023 \\
\hline
$\alpha=-1.0\chi$ & 1.399 & 1.470 & 2.020 & 2.026 \\
\hline
$\alpha=0$ (GR)   & 1.40 & 1.47 & 2.021 & 2.027 \\
\hline
$\alpha=+2.0\chi$ & 1.427 & 0.447 & 6.771 & 6.840 \\
\hline
$\alpha=+5.0\chi$ & 1.537 & 0.432 & 7.553 & 7.639 \\
\hline
\end{tabular}
\end{table}


\newpage

\section{Conclusions}\label{Sec:VII}

Observations suggest that super-Chandrasekhar white dwarfs may exist, implying that the classical Chandrasekhar limit can be surpassed. This possibility has been investigated in the context of modified theories of gravity \citep{das_2015, Otoniel_2026, Li_2024, banerjee_2017, Rocha_2020}. In this paper, we explore matter geometry coupled gravity, such as $f(\mathcal{R}, \mathcal{L}_m)$ and $f(\mathcal{R,L}_m, \mathcal{T})$, with specific functional forms. The generalized TOV equations are solved for the functional forms of each gravity using the Chandrasekhar EoS. Further, we vary the parameters and obtain $\rm M-\rm R$ relations for each functional form of $f(\mathcal{R}, \mathcal{L}_m)$ and $f(\mathcal{R,L}_m, \mathcal{T})$ gravity.

To assess the physical viability of the obtained configurations, we examined several stability criteria. For the power-law model of $f(\mathcal{R,L}_m)$ gravity, the effect of the coupling parameter $\sigma$ deviates from GR, and higher maximum masses are obtained for negative values of the parameter. With the values of $\sigma= -1.5\eta$ the  model provide $1.5 \mathrm{M}_\odot$  with radius about $0.439 \times 10^{-3} \,\mathrm{R}_\odot$ (Figure~\ref{fig:MR_f(R,Lm)_M1}). Similar results are observed in the non-linear form of $f(\mathcal{R,L}_m, \mathcal{T})$ gravity. With $\alpha \mathcal{RT}$  coupling term, the maximum mass exceeds the Chandrasekhar limit and increases with the increase of $\alpha$. For $\alpha=+5.0\chi$, $f(\mathcal{R,L}_m, \mathcal{T})$ gravity with non-linear case provided the maximum mass of $1.537\, \mathrm{M}_\odot$ and radius of $0.432\times 10^{-3}\, \mathrm{R}_\odot$ (Figure~\ref{fig:MR_f(R,Lm,T)_M1}).  Finally, we consider the square-root exponential model in the framework of the $f(\mathcal{R,L}_m)$ gravity. In this form, the deviation in masses is clearly observed and the maximum mass exceeds the Chandrasekhar limit and increases with increasing $\gamma$. For $\gamma=+5.5 \zeta$, the model obtain maximum mass up to $2.08\, \mathrm{M}_\odot$ with the radius of $1.72\times10^{-3}\, \mathrm{R}_\odot$ (Figure~ \ref{fig:MR_f(R,Lm)_M2}). We have analyzed the static stability criterion where $d\rm M/d \rho_c > 0$. All the functional forms of each gravity show an increase in mass with increasing central density up to a maximum mass, indicating consistency with the hydrostatic stability criterion (Figures \ref{fig:MRho_f(R,Lm)_M1} and \ref{fig:MRho_f(R,Lm,T)_M1}). Also, we studied compactness and gravitational redshift and find that, in each model of $f(\mathcal{R,L}_m)$ and $f(\mathcal{R,L}_m, \mathcal{T})$ gravity, the values are of order $\mathcal{O}(10^{-3})$, which is very small compared to those of NSs and BHs.  Furthermore, we evaluated the adiabatic index throughout the stellar interior and found that $\Gamma>4/3$ for all the functional forms of $f(\mathcal{R,L}_m)$ and $f(\mathcal{R,L}_m, \mathcal{T})$ gravity as shown in the Figures \ref{fig:Adiabatic_f(R,Lm)_M1} and \ref{fig:Adiabatic_f(R,Lm,T)_M1}. These results indicate that the obtained WD solutions are consistent with the hydrostatic and adiabatic stability criteria considered in this work. However, a definitive assessment of dynamical stability would require a dedicated radial perturbation analysis within the corresponding gravity framework.

Overall, our results suggest that matter--curvature coupled gravity can support WD sequences with a maximum mass limit that differs from the existing limit in the Newtonian scenario.  The resulting configurations satisfy the hydrostatic and adiabatic stability criteria examined in this work. Future studies may extend the present analysis by incorporating strong magnetic fields through the Einstein-Maxwell system, adopting more realistic multi-component thermal EoS, and performing a full radial perturbation analysis to investigate dynamical stability in these modified gravity theories. Further, one can analyze with strong magnetic fields, i.e., coupled with the Maxwell equations, and also consider a multi-component thermal EoS composed of He-C-O to examine the effects on the structure properties. 

\section*{Acknowledgement}
The author SKM acknowledges that the Ministry of Higher Education, Research, and Innovation (MoHERI) supported this research work through the project Grant No. BFP/RGP/CBS/24/203. NP acknowledges the financial support provided by the University Grants Commission (UGC) through the Junior Research Fellowship to carry out the research work. KNS and BM acknowledge the support of IUCAA, Pune (India), through the visiting associateship program. The SKM is also grateful to the UoN administration for its continued support and encouragement of the research work.

\bibliography{References}{}
\bibliographystyle{aasjournalv7}

\end{document}